%% file: paper.tex
\documentclass[orivec]{llncs}

\input{preamble}
\iflong
\fi

\begin{document}

\title{Spatial Interpolants}
\author{Aws Albargouthi\inst{1} \and Josh Berdine\inst{2} \and Byron Cook\inst{3} \and Zachary Kincaid\inst{4}}
\institute{University of Wisconsin-Madison$^1$, Microsoft Research$^2$, University College London$^3$, University of Toronto$^4$}

\maketitle
\begin{abstract}

We propose \alg, a new technique for proving 
properties of heap-manipulating programs that marries 
(1) a new \emph{separation logic--based} analysis for heap reasoning with 
(2) an \emph{interpolation-based} technique for refining 
heap-shape invariants with data invariants. 
\alg is \emph{property directed}, \emph{precise}, and produces counterexample 
traces when a property does not hold.  
Using the novel notion of \emph{spatial interpolants modulo theories}, 
\alg can infer complex invariants over general recursive predicates, 
e.g., of the form \emph{all elements in a linked list are even}
or a \emph{binary tree is sorted}. Furthermore, we treat interpolation as a black box, 
which gives us the freedom to encode data manipulation in any
suitable theory for a given program (e.g., bit vectors, arrays, or linear arithmetic), 
so that our technique immediately benefits from 
any future advances in SMT solving and interpolation.
\end{abstract}

\input{intro2}

\input{example}

\input{prelims}

\input{spint}

\input{snint}

\input{bounded_abduction}

\input{eval}

\input{related}

\bibliographystyle{splncs03}
\bibliography{paper}

\iflong
\clearpage
\appendix
\newpage
\input{alg}

\input{appendix}
\fi

\end{document}

%% file: preamble.tex
\usepackage{amsmath,amssymb,stmaryrd}
\usepackage{array}
\usepackage{mathpartir}
\usepackage{listings}
\usepackage[usenames,dvipsnames]{xcolor}
\usepackage{graphicx}
\usepackage{latexsym}
\usepackage{dsfont}
\usepackage[sort&compress,sectionbib,numbers,square]{natbib}
\usepackage{fix-cm}
\renewcommand{\bibfont}{\fontsize{8}{10}\selectfont}

\let\llncssubparagraph\subparagraph
\let\subparagraph\paragraph
\usepackage[small,compact]{titlesec}
\let\subparagraph\llncssubparagraph

\usepackage[noend]{algpseudocode}
\usepackage{algorithm}

\usepackage{xparse}
\newcommand{\figsep}[1]{%
  \null\hfill\rule[0.5ex]{2cm}{0.75pt} \small{{\textbf{\emph{#1}}}} \rule[0.5ex]{2cm}{0.75pt}\hfill\null\par%
}

\usepackage{xspace}
\usepackage{url}
\usepackage{wrapfig}

\usepackage{paralist}

\usepackage{setspace}

\renewcommand{\triangleright}{\blacktriangleright}
\newcommand{\impact}{\textsc{Impact}\xspace}

\newcommand{\lda}[1]{(\lambda #1)}

\newcommand{\ls}{\textsf{ls}}
\newcommand{\bt}{\textsf{bt}}
\newcommand{\nil}{\textsf{null}}
\newcommand{\true}{\textsf{true}}
\newcommand{\etrue}{\emph{true}}
\newcommand{\ltrue}{\emph{true}}
\newcommand{\emp}{\textsf{emp}}
\newcommand{\hoare}[3]{\{#1\}\;\texttt{#2}\;\{#3\}}
\newcommand{\sem}[1]{\llbracket #1 \rrbracket}
\newcommand{\tuple}[1]{\langle #1 \rangle}
\renewcommand{\phi}{\varphi}

\newcommand{\dom}{\text{dom}}

\newcommand{\domain}{\mathds{D}}

\newcommand{\logic}{\textsf{RSep}\xspace}
\newcommand{\sep}{\textsf{Sep}\xspace}
\newcommand{\Atom}{H}
\newcommand{\rec}{\textsf{Rec}}
\newcommand{\prog}{\mathcal{P}}
\newcommand{\hvar}{\textsf{HVar}}

\newcommand{\Vars}{\textsf{Var}}
\newcommand{\dvar}{\textsf{DVar}}
\newcommand{\aterm}{\textsf{DTerm}\xspace}
\newcommand{\hterm}{\textsf{HTerm}\xspace}

\newcommand{\pred}{\textsf{RPred}\xspace}

\newcommand{\alg}{\textsc{SplInter}\xspace}
\newcommand{\theory}{\mathcal{T}}
\newcommand{\interp}[3]{\textsf{itp}(#1, #2, #3)}
\newcommand{\post}{\textsf{exec}}
\newcommand{\refine}{\textsf{Refine}}
\newcommand{\ppath}{\textsf{ProvePath}}
\newcommand{\conj}{\textsf{Conj}\xspace}
\newcommand{\cproof}{\textsf{IsProof}}
\newcommand{\false}{\emph{false}}

\newcommand{\qex}[2]{(\exists #1.\;#2)}
\newcommand{\qs}[3]{\qex{#1}{#2 : #3}}
\newcommand{\ul}[1]{\underline{#1}}
\newcommand{\ol}[1]{\overline{#1}}
\newcommand{\abduce}[4]{#1 \vdash \qex{#2}{#3 * [A]} \vdash #4}
\newcommand{\cjudge}[2]{#2 \ \triangleright\ #1}

\newcommand{\eoe}{
  \hspace*{\fill}\ensuremath{\lrcorner}%
}

\usepackage[scaled=0.80]{DejaVuSansMono}

\lstdefinestyle{customc}{
  belowcaptionskip=1\baselineskip,
  breaklines=true,
  xleftmargin=\parindent,
  language=C,
  showstringspaces=false,
  basicstyle=\footnotesize\ttfamily,
  keywordstyle=\ttfamily\color{blue},
  commentstyle=\itshape\color{black},
  identifierstyle=\ttfamily\color{black},
  stringstyle=\color{orange},
  morekeywords={assert,assume,node}
}

\newif\iflong
\longtrue
\newcommand{\ifshort}[1]{\iflong\else#1\fi}

\newcommand{\shortenpar}{
  \ifshort{\looseness=-1}%
}

\renewcommand{\leq}{\leqslant}
\renewcommand{\geq}{\geqslant}

\definecolor{dgray}{gray}{0.35}

%% file: intro2.tex
\section{Introduction} \label{sec:intro}

Since the problem of determining whether a program satisfies a given property
is undecidable, every verification algorithm must make some compromise.  There
are two classical schools of program verification, which differ in
the compromise they make: the \emph{static analysis} school gives up
refutation soundness (i.e., may report \emph{false positives}); and the
\emph{software model checking} school gives up the guarantee of termination.
In the world of integer program verification, both schools are well explored
and enjoy
cross-fertilization of ideas:
each has its own strengths and uses in different contexts.
In the world of heap-manipulating programs, the
static analysis school is well-attended
\cite{Sagiv99,Distefano2006,Chang2008,Calcagno09}, while the software model
checking school has remained essentially vacant.  This paper initiates a
program to rectify this situation, by proposing one of the first path-based software model
checking algorithms for proving combined shape-and-data properties.

The algorithm we propose, \alg, marries two celebrated program
verification ideas: McMillan's \emph{lazy abstraction with interpolants} (\impact)
algorithm for software model checking \cite{McMillan2006}, and
\emph{separation logic}, a program logic for reasoning about shape properties
\cite{Reynolds2002}.  \alg (like \impact) is based on a path-sampling
methodology: given a program $P$ and safety property $\varphi$, \alg
constructs a proof that $P$ is memory safe and satisfies $\varphi$ by sampling
a finite number of paths through the control-flow graph of $P$, proving them
safe, and then assembling proofs for each sample path into a proof for the
whole program.  The key technical advance which enables \alg is an algorithm
for \emph{spatial interpolation}, which is used to construct proofs in
\emph{separation logic} for the sample traces (serving the same function as
\emph{Craig interpolation} for first-order logic in \impact).
\shortenpar

\alg is able to prove properties requiring integrated heap and data
(e.g., integer) reasoning by strengthening separation logic proofs with
\emph{data refinements} produced by classical Craig interpolation, using a
technique we call \emph{spatial interpolation modulo theories}.  Data
refinements are \emph{not tied to a specific logical theory}, giving us a
rather generic algorithm and freedom to choose an appropriate theory to encode
a program's data.

\begin{figure}[t]
\centering
\includegraphics[scale=0.32]{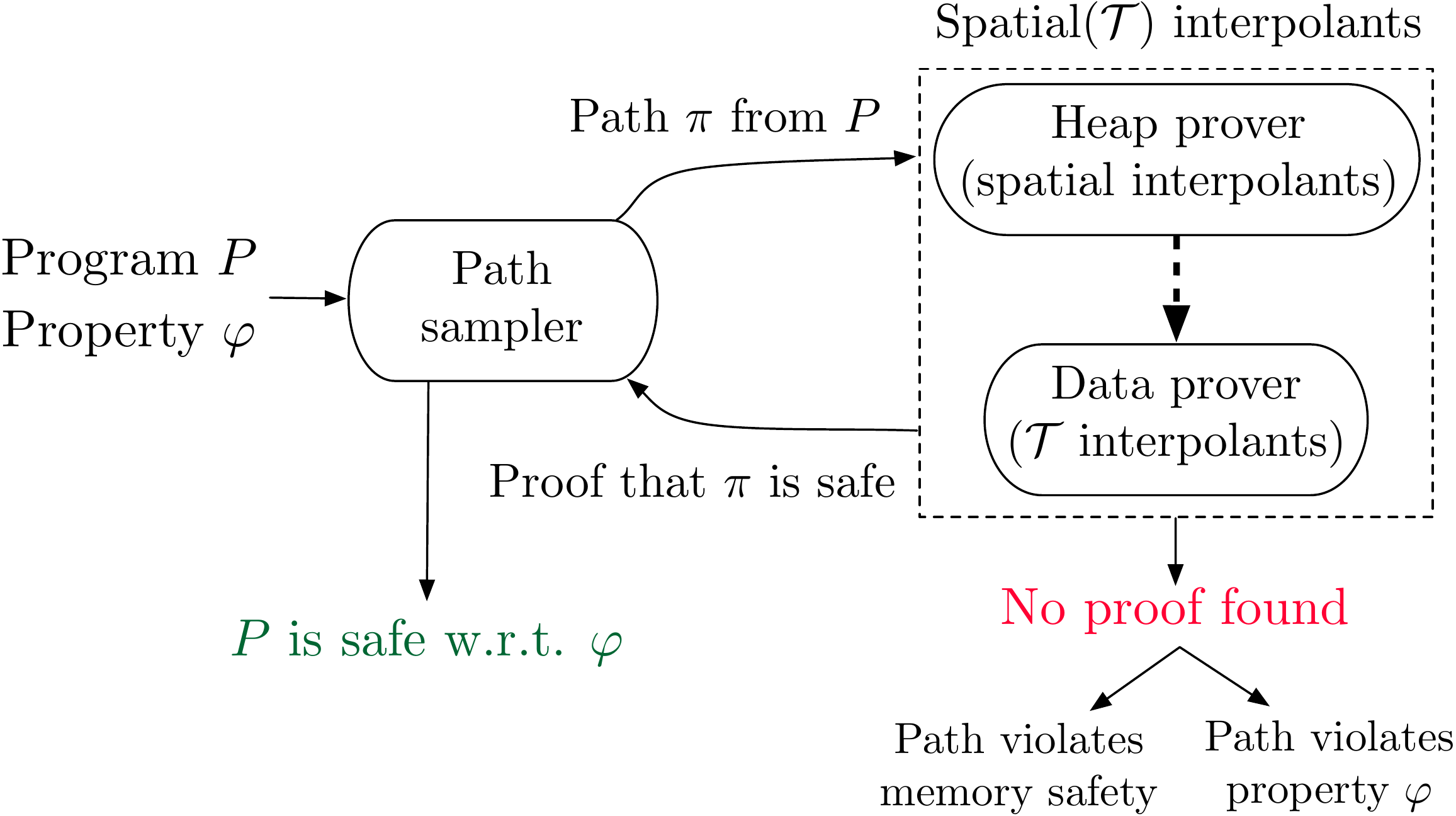}
\caption{Overview of \alg verification algorithm.}
\label{fig:arch}
\end{figure}

Fig.~\ref{fig:arch} summarizes the high-level
operation of our algorithm.
Given a program with no heap manipulation,
\alg only computes theory interpolants
and behaves exactly like
\impact, and thus
one can thus view \alg as a proper
extension of \impact to heap manipulating programs.
At the other extreme, given
a program with no data manipulation, \alg is a new shape analysis
that uses path-based relaxation to construct memory safety proofs in
separation logic.
\shortenpar

There is a great deal of work in the static analysis school on shape analysis
and on combined shape-and-data analysis, which we will discuss further in
Sec.~\ref{sec:rel}.  We do not claim superiority over these techniques
(which have had the benefit of 20 years of active development).  \alg, as the
first member of the software model checking school, is not \emph{better};
however, it \emph{is} fundamentally \emph{different}.  Nonetheless, we will
mention two of the features of \alg (not enjoyed by any previous verification
algorithm for shape-and-data properties) that make our approach worthy of
exploration: path-based refinement and property-direction.
\begin{compactitem}
\item \emph{Path-based refinement}:
  This supports a progress guarantee by tightly correlating program
  exploration with refinement, and by avoiding imprecision due to
  lossy join and widening operations employed by abstract domains.  \alg does
  not report false positives, and produces counterexamples for violated
  properties.
  This comes, as usual, at the price of potential divergence.
\item \emph{Property-direction}: Rather than seeking the strongest invariant
  possible, we compute one that is \emph{just strong enough} to prove that a
  desired property holds.  Property direction enables scalable reasoning in
  rich program logics like the one described in this paper, which combines
  separation logic with first-order data refinements.
\end{compactitem}

We have implemented an instantiation of our generic technique 
in the \textsc{T2} verification tool~\cite{t2},
and used it to prove correctness of a number of
programs, partly drawn from open source software,
requiring combined data and heap invariants.
Our results indicate the usability and promise
of our approach.

\paragraph{Contributions}
We summarize our contributions as follows:
\begin{enumerate}
\item A generic property-directed algorithm
for verifying and falsifying safety of programs with heap and data manipulation.
\item A precise and expressive separation logic analysis for computing memory safety
proofs of program paths using a novel technique we term \emph{spatial interpolation}.
\item A novel interpolation-based technique for strengthening
separation logic proofs with
data refinements.
\item An implementation and an evaluation of our technique for a fragment
of separation logic with linked lists enriched with linear arithmetic
refinements.
\end{enumerate}

%% file: example.tex
\section{Overview} \label{sec:ex}

In this section, we demonstrate the operation
of \alg (Fig.~\ref{fig:arch}) on the simple linked list example shown in Fig.~\ref{code:ex}.
We assume that integers are unbounded
(i.e., integer values are drawn from $\mathbb{Z}$ rather than machine integers) and%
\setlength{\intextsep}{0pt}%
\setlength{\columnsep}{5pt}%
\begin{wrapfigure}{r}{4.6cm}
\hfill
\begin{minipage}{4.6cm}
\begin{lstlisting}
1: int i = nondet(); 
   node* x = null;
2: while (i != 0)
     node* tmp = malloc(node);
     tmp->N = x;
     tmp->D = i;
     x = tmp;
     i--;
3: while (x != null)
4:   assert(x->D >= 0);
     x = x->N;
\end{lstlisting}
\end{minipage}
\caption{Illustrative Example \label{code:ex}}
\end{wrapfigure}%
that there is a \texttt{struct} called
\texttt{node} denoting a linked list node,
with a next  pointer \texttt{N} and an integer (data) element \texttt{D}. The function
\texttt{nondet()} returns a nondeterministic integer value.
This program starts by building a linked list
in the loop on location \texttt{2}.
The loop terminates if the initial value of \texttt{i}
is $\geq 0$, in which case a linked list
of size \texttt{i} is constructed, where data elements \texttt{D} 
of list nodes range from \texttt{1} to \texttt{i}.
Then, the loop at location \texttt{3} iterates
through the linked list asserting that
the data element of each node in the list is $\geq 0$.
Our goal is to prove that the assertion at location \texttt{4}
is never violated.

\begin{figure}[t!]
\centering
\includegraphics[width=\textwidth]{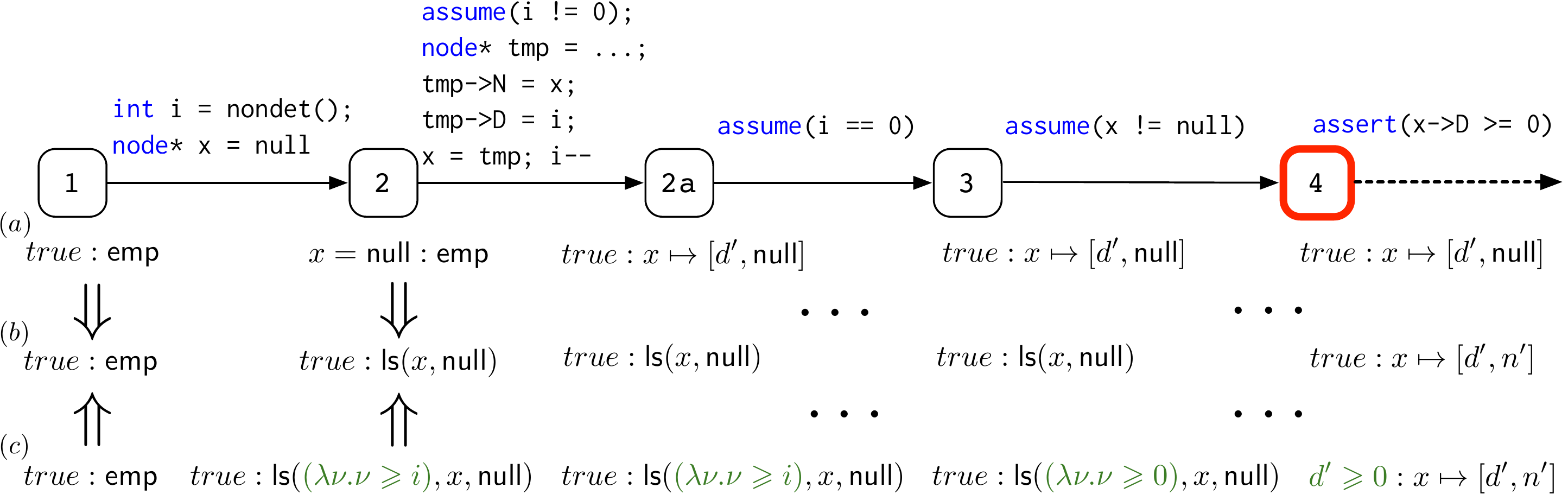}
\caption{
Path through program in Fig.~\ref{code:ex}, annotated with 
(a) results of forward symbolic execution,
(b) spatial interpolants,
and (c)
spatial($\theory$) interpolants,
where $\theory$ is linear integer arithmetic. Arrows $\Rightarrow$ indicate implication (entailment) direction. 
}
\label{fig:ex}
\end{figure}

\paragraph{Sample a Program Path}
To start, we need
a path $\pi$
through the program to the assertion at location \texttt{4}.
Suppose we start by sampling
the path \texttt{1,2,2,3,4},
that is, the path that goes through the first loop once,
and enters the second loop arriving at the assertion.
This path is illustrated in Fig.~\ref{fig:ex}
(where \texttt{2a}
indicates the second occurrence
of location \texttt{2}).
Our goal is to construct a Hoare-style
proof of this path: an annotation of each location
along the path with a formula describing
reachable states, such that location \texttt{4}
is annotated with a formula implying
that \texttt{x->D >= 0}.
This goal is accomplished in two phases. First, we use \emph{spatial interpolation} to compute a memory safety proof for the path $\pi$ (Fig.~\ref{fig:ex}(b)). Second,
we use \emph{theory refinement} to strengthen the memory safety proof and
establish that the path satisfies the post-condition
\texttt{x->D >= 0} (Fig.~\ref{fig:ex}(c)).

\paragraph{Compute Spatial Interpolants}
The first step in constructing
the proof is to find \emph{spatial interpolants}:
a sequence of separation logic formulas \emph{approximating}
the shape of the heap at each program location,
and forming a Hoare-style memory safety proof of the path.
Our spatial interpolation procedure is a two step process that first
symbolically executes the path in a forward pass and then derives a weaker
proof using a backward pass.  The backward pass can be thought of as an
under-approximate weakest precondition computation, which uses the symbolic
heap from the forward pass to guide the under-approximation.

We start by showing
the \emph{symbolic heaps} in Fig.~\ref{fig:ex}(a), which are the result of the forward pass obtained by
symbolically executing \emph{only} heap statements along
this program path (i.e., the strongest postcondition
along the path).
The separation logic annotations in Fig.~\ref{fig:ex}
follow standard notation (e.g.,~\cite{Distefano2006}),
where a formula is of the form $\Pi:\Sigma$,
where $\Pi$ is a Boolean first-order formula over heap variables (pointers)
as well as data variables (e.g., $x = \nil$ or $i > 0$),
and $\Sigma$ is a \emph{spatial conjunction}
of \emph{heaplets} (e.g., \textsf{emp}, denoting the empty heap,
or $Z(x,y)$, a recursive predicate, e.g., that denotes
a linked list between $x$ and $y$).
For the purposes of this example,
we assume a recursive predicate $\ls(x,y)$ that describes linked lists.
In our example, the symbolic heap
at location \texttt{2a} is $true:x\mapsto[d',\nil],$
where the heap consists of a node, pointed to by variable $x$, 
with $\nil$ in the \texttt{N} field and the (implicitly existentially
quantified) variable $d'$ in the \texttt{D} field (since so far we are only
interested in heap shape and not data).
\shortenpar

The symbolic heaps determine a memory safety proof of the path,
but it is too strong and would likely not generalize to other paths.
The goal of spatial interpolation is to find a sequence
of annotations that are weaker than the symbolic heaps,
but that still prove memory safety of the path.
A sequence of spatial interpolants is shown in Fig.~\ref{fig:ex}(b).
Note that all spatial interpolants are implicitly spatially conjoined
with $\true$; for clarity, we avoid explicitly conjoining formulas with
$\true$ in the figure.
For example, location \texttt{2} is annotated with $true : \ls(x,\nil) * \true$,
indicating that there is a list on the heap, as well as other potential objects
not required to show memory safety.
We compute spatial interpolants by going backwards along the path
and asking questions of the form: \emph{how much can we weaken
the symbolic heap while still maintaining memory safety?}
We will describe how to answer such questions
in Section~\ref{sec:spint}.

\paragraph{Refine with Theory Interpolants}
Spatial interpolants give us a memory safety proof
as an approximate heap shape at each location.
Our goal now is to strengthen these heap shapes
with data refinements, in order to prove that
the assertion at the end of the path is not violated.
To do so, we generate a system of Horn clause constraints from the path in some
first-order theory admitting interpolation (e.g., linear arithmetic).
These Horn clauses carefully encode
the path's data manipulation along with the spatial interpolants,
which tell us heap shape at each location along the path.
A solution of this constraint system, which can be solved
using off-the-shelf
interpolant generation techniques (e.g.,~\cite{McMillan2011,Rybalchenko07}),
is a \emph{refinement} (strengthening) of the memory safety proof.
\shortenpar

In this example, we encode program operations over integers
in the theory of linear integer arithmetic, and use Craig interpolants
to solve the system of constraints.
A solution of this system is a set of linear arithmetic
formulas that refine our spatial interpolants and,
as a result, imply the assertion we want to prove holds.
One possible solution is shown in Fig.~\ref{fig:ex}(c).
For example, location \texttt{2a} is now labeled
with $true \color{black} : 
\mathsf{ls}(\color{OliveGreen} \lda{\nu. \nu \geq i} \color{black} ,x,\mathsf{null}),$
where the {\color{OliveGreen}green} parts of the formula are those added by refinement.
Specifically, after refinement, we know that \emph{all} elements in the 
list from $x$ to $\nil$ after the first loop 
have data values greater than or equal to $i$,
as indicated by the predicate $\color{OliveGreen} \lda{\nu. \nu \geq i}$.
(In Section~\ref{sec:prelims}, we formalize recursive predicates
with data refinements.)

Location \texttt{4} is now annotated with 
$\color{OliveGreen}d' \geq 0 \color{black} : x \mapsto [d',n'] * \true$,
which implies that \texttt{x->D >= 0}, thus proving that the path
satisfies the assertion.

\paragraph{From Proofs of Paths to Proofs of Programs}
We go from proofs of paths to whole program proofs
implicitly by building an \emph{abstract reachability tree}
as in \textsc{Impact}~\cite{McMillan2006}.  To give a flavour for how this works, consider that the assertions at
\texttt{2} and \texttt{2a} are identical: this implies that this assertion is an
inductive invariant at line \texttt{2}.  Since this assertion also happens to be
strong enough to prove safety of the program, we need not sample any longer unrollings of the first loop. However, since we have not established the inductiveness of the assertion at \texttt{3}, the proof is not yet complete and more traces need to be explored (in fact, exploring one more trace will do: consider the trace that unrolls the second loop once and shows that the second time \texttt{3} is visited can also be labeled with $\color{black} true \color{black} : \mathsf{ls}(\color{OliveGreen} \lda{\nu. \nu \geq 0} \color{black} ,x,\mathsf{null})$).
\shortenpar

Since our high-level algorithm is virtually the same as
\textsc{Impact}~\cite{McMillan2006}, we will not describe it further in the
paper.
For the remainder of this paper, we will concentrate on the novel
contribution of our algorithm: computing spatial interpolants with
theory refinements for program paths.

%% file: prelims.tex
\section{Preliminaries} \label{sec:prelims}

\subsection{Separation Logic}

We define \logic, a fragment of separation logic formulas featuring points-to predicates and general recursive predicates refined by theory propositions.

\begin{figure}[t]
\centering
\fontsize{8.4}{9}\selectfont
$
\begin{array}{lll||lrl}
    x,y \in \textsf{HVar}& \hspace*{10pt}\text{(Heap variables)} & 
  && E,F \in \textsf{HTerm}&::= \nil \mid x\\

  a,b \in \dvar& \hspace*{10pt}\text{(Data variables)} &
  && \AE                  &::= A \ \mid E\\

  A \in \textsf{DTerm} & \hspace*{10pt} \text{(Data terms)} &
  && \Pi \in \textsf{Pure}&::= \etrue \mid E = E \mid E \neq E \mid\\

    \phi \in \textsf{DFormula} & \hspace*{10pt} \text{(Data formulas)}
    && &&~~~~~\varphi \ \mid \Pi \land \Pi\\

  Z \in \pred & \hspace*{10pt} \text{(Rec. predicates)} & 
  && \Atom \in \textsf{Heaplet}&::= \true \mid \emp \mid E \mapsto [\vec{A},\vec{E}] \mid Z(\vec{\theta},\vec{E})\\

  \theta \in \textsf{Refinement} &::= \lambda \vec{a}. \phi & 
  && \Sigma \in \textsf{Spatial}&::= \Atom \mid \Atom * \Sigma\\

  X \subseteq \textsf{Var}&::= x \mid a &
  && P \in \logic&::= \qs{X}{\Pi}{\Sigma}
\end{array}
$
\caption{Syntax of \logic formulas.}
\label{fig:logic-syntax}
\end{figure}

Fig.~\ref{fig:logic-syntax} defines the syntax
of \logic formulas.
In comparison with the standard list fragment 
used in separation logic analyses (e.g.,~\cite{Berdine2005,Cook2011,Perez2011}), 
the  differentiating features of \logic  are:
(1) General \emph{recursive predicates},
for describing unbounded pointer structures like 
lists, trees, etc.
(2) Recursive predicates are augmented with 
a vector of \emph{refinements}, which are used to constrain
the data values appearing on the data structure defined by the predicate, 
detailed below.
(3) Each heap cell (points-to predicate), $E \mapsto [\vec{A},\vec{E}]$, 
is a \emph{record} consisting of \emph{data} fields (a vector $\vec{A}$ of \aterm)
followed by \emph{heap} fields (a vector $\vec{E}$ of \hterm).
(Notationally, we will use $d_i$ to refer to the $i$th element of the vector
$\vec{d}$, and $\vec{d}[t/d_i]$ to refer to the vector $\vec{d}$ with the
$i$th element modified to $t$.)
(4) \textsf{Pure} formulas
contain heap and
first-order data constraints.

Our definition is (implicitly) parameterized by a first-order theory
$\theory$.  $\dvar$ denotes the set of theory variables,
which we assume to be disjoint from $\hvar$ (the set of heap variables).
$\textsf{DTerm}$ and $\textsf{DFormula}$ denote the sets of theory terms and formulas, and we assume
that heap variables do not appear in theory terms.

For an \logic formula $P$,
$\Vars(P)$
denotes its free (data and heap)
variables.
We
treat a $\textsf{Spatial}$ formula $\Sigma$
as a multiset of heaplets, and consider formulas to be equal when they are equal as multisets.  For \logic formulas $P =
\qs{X_P}{\Pi_P}{\Sigma_P}$ and $Q = \qs{X_Q}{\Pi_Q}{\Sigma_Q}$, we
write $P*Q$ to denote the \logic formula
\shortenpar
\[ P * Q = \qs{X_P \cup X_Q}{\Pi_P \land \Pi_Q}{\Sigma_P * \Sigma_Q}  \]
assuming that $X_P$ is disjoint from $\Vars(Q)$ and $X_Q$ is disjoint from $\Vars(P)$
(if not, then $X_P$ and $X_Q$ are first suitably
renamed).  For a set of variables $X$, we write $\qex{X}{P}$ to denote the
\logic formula
\[ \qex{X}{P} = \qs{X \cup X_P}{\Pi_P}{\Sigma_P} \]

\paragraph{Recursive predicates}
Each recursive predicate $Z \in \pred$ is associated with a definition that
describes how the predicate is unfolded.  Before we formalize these
definitions, we will give some examples.
\shortenpar

The definition of the list segment predicate from Sec.~\ref{sec:ex} is:
\[\begin{split}
\ls(R,x,y) \equiv {}& (x = y : \emp) \lor {}\\
& \qs{d,n'} {x \neq y \land R(d)} {x \mapsto [d,n'] * \ls(R,n',y)}
\end{split}\]
In the above, $R$ is a \emph{refinement variable}, which may be instantiated
to a concrete refinement $\theta \in \textsf{Refinement}$.  For example,
$\ls(\lda{a . a \geq 0},x,y)$ indicates that there is a list from $x$ to
$y$ where every element of the list is at least $0$.

A refined binary tree predicate is a more complicated example:
\begin{align*}
\textsf{bt}(Q,L,R,x) =&\; (x = \nil : \emp)\\
\lor &\; \qex{d,l,r}{Q(d) :  x \mapsto [d,l,r] \\
&\hspace*{0.5cm} * \textsf{bt}((\lambda a. Q(a) \land L(d,a)), L, R, l) \\
&\hspace*{0.5cm} * \textsf{bt}((\lambda a. Q(a) \land R(d,a)), L, R, r)}
\end{align*}
This predicate has three refinement variables: a unary refinement $Q$ (which
must be satisfied by every node in the tree), a binary refinement $L$ (which is a
relation that must hold between every node and its descendants to the left),
and a binary refinement $R$ (which is a relation that must hold between every
node and its descendants to the right).  For example,
\[ \textsf{bt}((\lambda a. \ltrue), (\lambda a, b. a \geq b), (\lambda a, b. a \leq b), x) \]
indicates that $x$ is the root of a \emph{binary search tree}, and
\[ \textsf{bt}((\lambda a. a \geq 0), (\lambda a, b. a \leq b), (\lambda a, b. a \leq b), x) \]
indicates that $x$ is the root of a \emph{binary min-heap} with non-negative
elements.

To formalize these definitions, we first define \emph{refinement terms} and
\emph{refined formulas}: a refinement term $\tau$ is either (1) a refinement
variable $R$ or (2) an abstraction $(\lambda a_1,\dotsc,a_n. \Phi)$, where $\Phi$
is a refined formula.  A \emph{refined formula} is a conjunction where each
conjunct is either a data formula (\textsf{DFormula}) or the application $\tau(\vec{A})$ of a
refinement term to a vector of data terms (\textsf{DTerm}).

A \emph{predicate definition} has the form
\[Z(\vec{R}, \vec{x}) \equiv \qex{X_1}{\Pi_1 \land \Phi_1 : \Sigma_1} \lor \dotsb \lor \qex{X_n}{\Pi_n \land \Phi_n : \Sigma_n}\]
where $\vec{R}$ is a vector of refinement variables, $\vec{x}$ is a vector of
heap variables, and where refinement terms may appear as
refinements in the spatial formulas $\Sigma_i$.
We refer to the disjuncts of the above formula as the \emph{cases} for $Z$,
and define $\mathit{cases}(Z(\vec{R},\vec{x}))$ to be the set of cases of $Z$.
$\vec{R}$ and $\vec{x}$ are bound in $\mathit{cases}(Z(\vec{R},\vec{x}))$, and we will assume that
predicate definitions are closed, that is, for each case of $Z$, the free
refinement variables belong to $\vec{R}$, the free heap variables belong to
$\vec{x}$, and there are no free data variables.
We also assume that they are well-typed in the sense that each
refinement term $\tau$ is associated with an arity, and whenever
$\tau(\vec{A})$ appears in a definition, the length of $\vec{A}$ is the arity
of $\tau$.
\shortenpar

\paragraph{Semantics}
The semantics of our logic, defined by a satisfaction relation $s,h \models
Q$, is essentially standard.
Each predicate $Z \in \pred$ is defined to be
the least solution\footnote{Our definition
  does not preclude ill-founded predicates; such predicates are simply
  unsatisfiable, and do not affect the technical development in the rest of
  the paper.} to the following equivalence:
\[
  s,h  \models Z(\vec{\theta},\vec{E}) \iff \exists P \in \mathit{cases}(Z(\vec{R},\vec{x})).\ s,h \models P[\vec{\theta}/\vec{R} , \vec{E}/\vec{x} ]
\]
Note that when substituting a $\lambda$-abstraction for a refinement variable,
we implicitly $\beta$-reduce
resulting applications.  For example,
$R(b)[(\lambda a. a \geq 0)/R] = b \geq 0$.

Semantic entailment is denoted by $P \models Q$, and provable entailment by $P \vdash Q$.
When referring to a proof that $P \vdash Q$,
we will mean a sequent calculus proof.

\subsection{Programs}
A program $\prog$  is a tuple $\tuple{V,E,v_{\rm i},v_{\rm e}}$,
where 
\begin{compactitem}
\item $V$ is a set of control locations, with a distinguished \emph{entry}
  node $v_{\rm i} \in V$ and \emph{error} (exit) node $v_{\rm e} \in V$, and
\item $E \subseteq V \times V$ is a set of directed edges, where each $e \in E$ is
associated with a program command $e^{\rm c}$.
\end{compactitem}

We impose the restriction that all nodes $V \setminus \{v_{\rm i}\}$
are reachable from $v_{\rm i}$ via $E$, and all nodes can reach $v_{\rm e}$.
The syntax for program commands appears below.
Note that the allocation command
creates a record with $n$ data fields, $D_1,\ldots,D_n$, and
$m$ heap fields, $N_1,\ldots,N_m$.
To access the $i$th data field of a record pointed to by \texttt{x}, 
we use \texttt{x->D$_i$} (and similarly for heap fields).
We assume that programs are well-typed, but not necessarily
memory safe.
~\\~\\
\noindent
{\small
\begin{minipage}{0.3\linewidth}
  \textbf{Assignment}: \texttt{x := \AE}\\
  \textbf{Heap store}: \texttt{x->N$_i$ := E}\\
  \textbf{Heap load}: \texttt{y := x->N$_i$}
\end{minipage}
\hfill
\begin{minipage}{0.3\linewidth}
  \textbf{Assumption}: \texttt{assume($\Pi$)}\\
  \textbf{Data store}: \texttt{x->D$_i$ := A}\\
  \textbf{Data load}: \texttt{y := x->D$_i$}
\end{minipage}
\hfill
\begin{minipage}{0.33\linewidth}
\textbf{Allocation}: \texttt{x := new($n,m$)}\\
\textbf{Disposal}: \texttt{free(x)}\\
\end{minipage}}

\mbox{}\\\noindent
As is standard, we compile assert commands to reachability of $v_{\rm e}$.

%% file: spint.tex
\section{Spatial Interpolants} \label{sec:spint}
In this section, we first define the notion of spatial path interpolants,
 which serve as memory safety proofs of program paths.  We then describe a
technique for computing spatial path interpolants.  This algorithm has two
phases: the first is a (forwards) \emph{symbolic execution} phase, which
computes the strongest memory safety proof for a path; the second is a
(backwards) \emph{interpolation} phase, which weakens the proof so that it is more likely to generalize.

Spatial path interpolants are bounded from below by the strongest memory
safety proof, and (implicitly) from above by the weakest memory safety proof.
Prior to considering the generation of inductive invariants using spatial path
interpolants, consider what could be done with only one of the bounds, in
general, with either a path-based approach or an iterative fixed-point
computation.  Without the upper bound, an interpolant or invariant could be
computed using a standard forward transformer and widening.  But this suffers
from the usual problem of potentially widening too aggressively to prove the
remainder of the path, necessitating the design of analyses which widen
conservatively at the price of computing unnecessarily strong proofs.  The
upper bound neatly captures the information that must be preserved for the
future execution to be proved safe.  On the other hand, without the lower
bound, an interpolant or invariant could be computed using a backward
transformer (and lower widening).  But this suffers from the usual problem
that backward transformers in shape analysis explode, due to issues such as
not knowing the aliasing relationship in the pre-state.  The lower bound
neatly captures such information, heavily reducing the potential for
explosion.  These advantages come at the price of operating over full paths
from entry to error.  Compared to a forwards iterative analysis, operating
over full paths has the advantage of having information about the execution's
past and future when weakening at each point along the path.  A forwards
iterative analysis, on the other hand, trades the information about the future
for information about many past executions through the use of join or widening
operations.

The development in this section is purely spatial: we do not make use of data
variables or refinements in recursive predicates.  Our algorithm is thus of
independent interest, outside of its context in this paper.  We use \sep to
refer to the fragment of \logic in which the only data formula (appearing in
pure assertions and in refinements) is $\textit{true}$ (this fragment is
equivalent to classical separation logic).
An \logic formula $P$, in particular including those in recursive predicate
definitions, determines a \sep formula $\ul{P}$ obtained by replacing all
refinements (both variables and $\lambda$-abstractions) with
$\lda{\vec{a}.\ltrue}$ and all $\textsf{DFormula}$s in the 
pure part of $P$ with $\ltrue$.  Since recursive predicates, refinements, and \textsf{DFormula}s 
appear only positively, $\ul{P}$ is no stronger than any refinement of $P$.
Since all refinements in $\sep$
are trivial, we will omit them from the syntax (e.g., we will write
$Z(\vec{E})$ rather than $Z((\lambda \vec{a}. \ltrue), \vec{E})$).

\subsection{Definition} \label{sec:spint_def}
\begin{figure}[t]
  {\footnotesize
  $\post(\texttt{x := new($k,l$)},\ \qs{X}{\Pi}{\Sigma}) =$  $ \qex{X \cup \{x',\vec{d},\vec{n}\}}{(\Pi : \Sigma)[x'/x] * x \mapsto [\vec{d},\vec{n}]}$\\
    \hspace*{0pt}\hfill where $x',\vec{d},\vec{n}$ are fresh, $\vec{d} = (d_1,\ldots,d_k)$, and $\vec{n} = (n_1,\ldots,n_l)$.

$\post(\texttt{free(x)},\ \qs{X}{\Pi}{\Sigma*z \mapsto [\vec{d},\vec{n}]} = \qs{X}{\Pi \land \Pi^{\neq}}{\Sigma}$\\
\hspace*{0pt}\hfill where $\Pi : \Sigma * z \mapsto [\vec{d},\vec{n}] \vdash x = z$ and
$\Pi^{\neq}$ is the \\
\hspace*{0pt}\hfill conjunction of all disequalities $x \neq y$ s.t $y \mapsto [\_, \_] \in \Sigma$.\vspace*{3pt}

  $\post(\texttt{x := E},\ \qs{X}{\Pi}{\Sigma}) =$ $\qex{X\cup\{x'\}}{(x = E[x'/x]) * (\Pi : \Sigma)[x'/x]}$\\
    \hspace*{0pt}\hfill where $x'$ is fresh. \vspace*{3pt}

  $\post(\texttt{assume}(\Pi'),\ \qs{X}{\Pi}{\Sigma}) = \qs{X}{\Pi \land \ul{\Pi'}}{\Sigma}$ . \vspace*{3pt}

  $\post(\texttt{x->N$_i$ := E},\ \qs{X}{\Pi}{\Sigma * z \mapsto [\vec{d},\vec{n}]}) =$ $\qs{X}{\Pi}{\Sigma * x \mapsto [\vec{d},\vec{n}[E/n_i]]}$\\
    \hspace*{0pt}\hfill where $i \leq |\vec{n}|$ and 
            $\Pi:\Sigma * z \mapsto [\vec{d},\vec{n}] \vdash x = z$ . \vspace*{3pt}

  $\post(\texttt{y := x->N$_i$},\ \qs{X}{\Pi}{\Sigma * z \mapsto [\vec{d},\vec{n}]}) =$\\ 
  \hspace*{0pt}\hfill$\qex{X\cup\{y'\}}{(y = n_i[y'/y]) * (\Pi : \Sigma * z \mapsto [\vec{d},\vec{n}])[y'/y]}$\\
    \hspace*{0pt}\hfill where $i \leq |\vec{n}|$ and 
    $\Pi:\Sigma * z \mapsto [\vec{d},\vec{n}] \vdash x = z$, and $y'$ is fresh.}
  \caption{Symbolic execution for heap statements.  Data statements are treated as skips.
  }
  \label{fig:post}
\end{figure}

We define a \emph{symbolic heap} to be a \sep formula where the spatial part
is a *-conjunction of points-to heaplets and the pure part is a conjunction of
pointer (dis)equalities.  Given a command $\texttt{c}$ and a symbolic heap
$S$, we use $\post(\texttt{c}, S)$ to denote the symbolic heap that results
from symbolically executing $\texttt{c}$ starting in $S$ (the definition of
$\post$ is essentially standard \cite{Berdine2005}, and is shown in 
Fig.~\ref{fig:post}).

Given a program path $\pi = e_1,\ldots,e_n$, we obtain its strongest memory
safety proof by symbolically executing $\pi$ starting from the empty heap
\emp.  We call this sequence of symbolic heaps the symbolic execution sequence
of $\pi$, and say that a path $\pi$ is \emph{memory-feasible} if every formula in
its symbolic execution sequence is consistent.  The following proposition
justifies calling this sequence the strongest memory safety proof.
\begin{proposition} \label{prop:sp}
  For a
  path $\pi$, if the symbolic execution sequence for $\pi$ is
  defined, then $\pi$ is memory safe.  If $\pi$ is memory safe and memory-feasible,
  then its symbolic execution sequence is defined.
\end{proposition}

Recall that our strategy for proving program correctness is based on sampling
and proving the correctness of several program paths (\emph{\'{a} la} {\sc
  Impact}~\cite{McMillan2006}).  The problem with \emph{strongest} memory
safety proofs is that they do not generalize well (i.e., do not generate inductive
invariants).

One solution to this problem is to take advantage of property direction.
Given a desired postcondition $P$ and a (memory-safe and -feasible) path $\pi$, the goal is
to come up with a proof that is weaker than $\pi$'s symbolic execution
sequence, but still strong enough to show that $P$ holds after executing
$\pi$.  Coming up with such ``weak'' proofs is how traditional path
interpolation is used in {\sc Impact}.  In light of this, we define
\emph{spatial path interpolants} as follows:
\begin{definition}[Spatial path interpolant]
  Let $\pi = e_1,\ldots, e_n$ be a program path with symbolic execution sequence
  $S_0,\ldots,S_n$, and let $P$ be a \sep formula (such
  that $S_n \models P$).  A \emph{spatial path interpolant} for $\pi$ is a
  sequence $I_0,\ldots,I_n$ of \sep formulas such that
  \shortenpar
  \begin{compactitem}
  \item for each $i \in [0,n]$, $S_i \models I_i$;
  \item for each $i \in [1,n]$, $\{I_{i-1}\}\;e^{\rm c}_i\;\{I_i\}$ is a valid
    triple in separation logic; and
  \item $I_n \models P$ .
  \end{compactitem}
\end{definition}

Our algorithm for computing spatial path interpolants is a backwards
propagation algorithm that employs a \emph{spatial
  interpolation} procedure at each backwards step.  Spatial
interpolants for a single command are defined as:
\begin{definition}[Spatial interpolant] \label{def:spint}
  Given \sep formulas $S$ and $I'$ and a command $\texttt{c}$ such that $\post(\texttt{c},S)
  \models I'$, a \emph{spatial interpolant} (for $S$, $\texttt{c}$, and $I'$) is a \sep formula
  $I$ such that $S \models I$ and $\hoare{I}{c}{I'}$ is valid.
\end{definition}

Before describing the spatial interpolation algorithm, we briefly describe how spatial interpolation is used
to compute path interpolants.  Let us use $\interp{S}{\texttt{c}}{I}$ to denote a
spatial interpolant for $S,\texttt{c},I$, as defined above.  Let $\pi = e_1,\ldots,e_n$ be
a program path and let $P$ be a \sep formula.  First, symbolically execute
$\pi$ to compute a sequence $S_0,\ldots,S_n$.
Suppose that $S_n \vdash P$.  Then we compute a sequence $I_0,\ldots,I_n$ by taking $I_n = P$ and (for $k < n$) $I_{k} = \interp{S_{k}}{e_{k+1}^{\rm c}}{I_{k+1}}$.
The sequence $I_0,\ldots,I_n$ is clearly a spatial path interpolant.
\shortenpar

\subsection{Bounded Abduction}

Our algorithm for spatial interpolation is based on an abduction
procedure.  Abduction refers to the inference of explanatory hypotheses from
observations (in contrast to deduction, which derives conclusions from given
hypotheses).
The variant of abduction we employ in this paper, which we call \emph{bounded
  abduction}, is simultaneously a form of abductive and deductive reasoning.
Seen as a variant of abduction, bounded abduction adds a constraint that the
abduced hypothesis be at least weak enough to be derivable from a given hypothesis.  Seen
as a variant of deduction, bounded abduction adds a constraint that the deduced
conclusion be at least strong enough to imply some desired conclusion.
Formally, we define bounded abduction as follows:
\begin{definition}[Bounded abduction]
  Let $L,M,R$ be \sep formulas, and let $X$ be a set of variables.  A
  solution to the \emph{bounded abduction problem}
  \[ L \vdash \qex{X}{M * [\ ]} \vdash R \]
  is a \sep formula $A$ such that
  $ L \models \qex{X}{M * A} \models R$.
\end{definition}
Note how, in contrast to bi-abduction~\cite{Calcagno09} where a
solution is a pair of formulas, one constrained from above and one
from below, a solution to bounded abduction problems is a single
formula that is simultaneously constrained from above and below.  The
fixed lower and upper bounds in our formulation of abduction give
considerable guidance to solvers, in contrast to bi-abduction, where
the bounds are part of the solution.

Sec.~\ref{sec:abduction} presents our bounded abduction algorithm.  For
the remainder of this section, we will treat bounded abduction as a black box,
and use $L \vdash \qex{X}{M * [A]} \vdash R$ to denote that $A$ is a solution
to the bounded abduction problem.

\subsection{Computing Spatial Interpolants}

We now proceed to describe our algorithm for spatial
interpolation.  Given a command $\texttt{c}$ and \sep formulas $S$ and $I'$ such that
$\post(\texttt{c},S) \vdash I'$, this algorithm must compute a \sep formula
$\interp{S}{\texttt{c}}{I'}$ that satisfies the conditions of Definition~\ref{def:spint}.
Several examples illustrating this procedure are given in Fig.~\ref{fig:ex}.

This algorithm is defined by cases based on the command $\texttt{c}$.  We present the
cases for the spatial commands; the corresponding data commands are similar.

\paragraph{Allocate} Suppose \texttt{c} is \texttt{x := new($n,m$)}.  We take
$ \interp{S}{\texttt{c}}{I'} = \qex{x}{A}$,
where $A$ is obtained as a solution to $\abduce{\post(\texttt{c},S)}{\vec{a},\vec{z}}{x \mapsto
  [\vec{a},\vec{z}]}{I'}$, and $\vec{a}$ and $\vec{z}$ are vectors of fresh
variables of length $n$ and $m$, respectively.

\paragraph{Deallocate} Suppose \texttt{c} is \texttt{free(x)}.  We take
$\interp{S}{\texttt{c}}{I'} = \qex{\vec{a},\vec{z}}{I' * x \mapsto [\vec{a},\vec{z}]}$,
where $\vec{a}$ and $\vec{z}$ are vectors of fresh variables whose lengths are
determined by the unique heap cell which is allocated to $x$ in $S$.

\paragraph{Assignment} Suppose \texttt{c} is \texttt{x := E}.  We take
$\interp{S}{\texttt{c}}{I'} = I'[E/x]$.

\paragraph{Store} Suppose \texttt{c} is \texttt{x->N$_i$ := E}.  We take
$\interp{S}{\texttt{c}}{I'} = \qex{\vec{a},\vec{z}}{A * x \mapsto [\vec{a},\vec{z}]}$,
where $A$ is obtained as a solution to
$\post(\texttt{c},S) \vdash \qex{\vec{a},\vec{z}}{x \mapsto [\vec{a},\vec{z}[E/z_i]] * [A]} \vdash I'$
and where $\vec{a}$ and $\vec{z}$ are vectors of fresh variables whose lengths
are determined by the unique heap cell which is allocated to $x$ in $S$.

\begin{example}
Suppose that $S$ is $t \mapsto [4,y,\nil] * x \mapsto [2,\nil,\nil]$
where the cells have one data and two pointer fields,
\texttt{c} is \texttt{t->N$_0$ := x}, and $I'$ is $\textsf{bt}(t)$.  Then we can
compute
$\post(\texttt{c},S) = t \mapsto [4,x,\nil] * x \mapsto [2,\nil,\nil]$,
and then solve the bounded abduction problem
\[ \post(\texttt{c},S) \vdash \qex{a,z_1}{t \mapsto [a,x,z_1] * [\ ]} \vdash I'\ .\]
One possible solution is $A = \bt(x) * \bt(z_1)$, which yields
\[
  \interp{S}{\texttt{c}}{I'} = \qex{a,z_0,z_1}{t \mapsto [a,z_0,z_1] * \bt(z_1) * \bt(x)}
  \ .\tag*{\ensuremath{\lrcorner}}
\]
\end{example}

\paragraph{Load} Suppose \texttt{c} is \texttt{y := x->N$_i$}.  
Suppose that
 $\vec{a}$ and $\vec{z}$ are vectors
  of fresh variables of lengths $|\vec{A}|$ and $|\vec{E}|$ where $S$ is of
  the form $\Pi : \Sigma * w \mapsto [\vec{A},\vec{E}]$ and $\Pi:\Sigma * w
  \mapsto [\vec{A},\vec{E}] \vdash x = w$ (this is the condition under which $\post(\texttt{c}, S)$ is defined, see Fig.~\ref{fig:post}).
Let $y'$ be a
fresh variable, and define $\ol{S} = (y = z_i[y'/y]) * (\Pi : \Sigma * w \mapsto
[\vec{a},\vec{z}])[y'/y]$.  Note that $\ol{S} \vdash \qex{y'}{\ol{S}} \equiv
\post(\texttt{c},S) \vdash I'$.

We take
$\interp{S}{\texttt{c}}{I'} = \qex{\vec{a},\vec{z}}{A[z_i/y,y/y'] * x \mapsto [\vec{a},\vec{z}]}$
where $A$ is obtained as a solution to $\ol{S} \vdash \qex{\vec{a},\vec{z}}{x[y'/y] \mapsto [\vec{a},\vec{z}] * [A]} \vdash I'$.

\begin{example}
Suppose that $S$ is $y = t : y \mapsto [1,\nil,x] * x \mapsto [5,\nil,\nil]$,
\texttt{c} is \texttt{y := y->N$_1$}, and $I'$ is $y \neq \nil : \bt(t)$.  
Then $\ol{S}$ is
\[
  y = x \land y' = t : y' \mapsto [1,\nil,x] * x \mapsto [5,\nil,\nil]
\]

We can then solve the bounded abduction problem
\[ \overline{S} \vdash \qex{a,z_0,z_1}{y' \mapsto [a,z_0,z_1] * [\ ]} \vdash I'\]
A possible solution is $y \neq \nil \land y' = t : \bt(z_0) * \bt(z_1)$, yielding\\
\hspace*{0.5cm}$\interp{S}{\texttt{c}}{I'} = (\exists a,z_0,z_1. z_1 \neq \nil \land y = t :\bt(z_0) * \bt(z_1) * y \mapsto[a,z_0,z_1])\ .$\hspace*{0.5cm}\null
\ensuremath{\lrcorner}
\end{example}

\paragraph{Assumptions} The interpolation rules defined up to this point cannot
introduce recursive predicates, in the sense that if $I'$ is a
\mbox{*-conjunction} of points-to predicates then so is
$\interp{S}{\texttt{c}}{I'}$.\footnote{But if $I'$ \emph{does} contain
  recursive predicates, then $\interp{S}{\texttt{c}}{I'}$ may also.}
A \mbox{*-conjunction} of points-to predicates is \emph{exact} in the sense
that it gives the full layout of some part of the heap.  The power of recursive 
predicates lies in their ability to be \emph{abstract} rather than
exact, and describe only the shape of the heap rather than its exact layout.
It is a special circumstance that $\hoare{P}{\texttt{c}}{I'}$
holds when $I'$ is exact in this sense and $P$ is not: intuitively, it means
that by executing \texttt{c} we somehow gain information about the program
state, which is precisely the case for \texttt{assume} commands.

For an example of how spatial interpolation can introduce a recursive predicate
at an \texttt{assume} command,
consider the problem of computing an
interpolant \[\interp{S}{\texttt{assume(x $\neq \nil$)}}{(\exists
  a,z.\; x\mapsto[a,z] * \true)}\] where $S \equiv x \mapsto [d,y] * y \mapsto
[d',\nil]$: a desirable interpolant may
be $\ls(x,\nil) * \true$.  
The disequality
introduced by the assumption ensures
that one of the $\mathit{cases}$ of
the recursive predicate $\ls(x,\nil)$ (where the list from $x$ to $\nil$ is
empty) is impossible, which implies that the other case (where $x$ is
allocated) must hold.

Towards this end, we now define an auxiliary function $\textsf{intro}$ which
we will use to introduce recursive predicates for the \texttt{assume}
interpolation rules.  Let $P,Q$ be \sep formulas such that $P \vdash Q$, let
$Z$ be a recursive predicate and $\vec{E}$ be a vector of heap terms.  We
define $\textsf{intro}(Z,\vec{E},P,Q)$ as follows: if
  $\abduce{P}{\emptyset}{Z(\vec{E})}{Q}$
  has a solution and $A \nvdash Q$, define
 $\textsf{intro}(Z,\vec{E},P,Q) = Z(\vec{E}) * A$.
  Otherwise, define
  $\textsf{intro}(Z,\vec{E},P,Q) = Q$.

  Intuitively, the abduction problem has a solution when $P$ implies
  $Z(\vec{E})$ and $Z(\vec{E})$ can be \emph{excised} from $Q$.  The condition $A
  \nvdash Q$ is used to ensure that the excision from $Q$ is non-trivial
  (i.e., the part of the heap that satisfies $Z(\vec{E})$ ``consumes'' some
  heaplet of $Q$).

  To define the interpolation rule for assumptions, suppose
  \texttt{c} is \texttt{assume(E $\neq$ F)} (the case of equality assumptions
  is similar).  Letting
  $\{\tuple{Z_i,\vec{E}_i}\}_{i\leq n}$ be an enumeration of the (finitely many)
  possible choices of $Z$ and $\vec{E}$, we define a formula $M$ to be the
  result of applying $\textsf{intro}$ to $I'$ over all possible choices of $Z$ and
  $\vec{E}$:
\shortenpar
  \[M = \textsf{intro}(Z_1,\vec{E}_1,S\land E \neq F,\textsf{intro}(Z_2,\vec{E}_2,S \land E \neq F,\dotsc))\]
  where the innermost occurrence of $\textsf{intro}$ in this definition is
  $\textsf{intro}(Z_n,\vec{E}_n,S\land E \neq F, I')$.
  Since \textsf{intro} preserves entailment (in the sense that if $P \vdash Q$
  then $P \vdash \textsf{intro}(Z,\vec{E},P,Q)$), we have that $S \land E \neq F
  \vdash M$.  From a proof of $S \land E \neq F \vdash M$, we can construct a
  formula $M'$ which is entailed by $S$ and differs from $M$ only in that it
  renames variables and exposes additional equalities and
  disequalities implied by $S$, and take $\interp{S}{\texttt{c}}{I'}$ to be this
  $M'$.

  The construction of $M'$ from $M$ is straightforward but tedious.
  \emph{The procedure is detailed in \iflong Appendix~\ref{sec:assum} \else the extended version\fi; here, we will just
  give an example to give intuition on why it is necessary.}
Suppose that $S$ is $x = w :
  y \mapsto z$ and $I'$ is $\ls(w,z)$, and \texttt{c} is \texttt{assume($x$ =
    $y$)}.  Since there is no opportunity to introduce new recursive
  predicates in $I'$, $M$ is simply $\ls(w,z)$.  However, $M$ is not a valid
  interpolant since $S \not\models M$, so we must expose the equality $x=w$
  and rename $w$ to $y$ in the list segment in $M' \equiv x = w: \ls(y,z)$.

  In practice, it is undesirable to enumerate all possible choices of $Z$ and
  $\vec{E}$ when constructing $M$ (considering that if there are $k$ in-scope
  data terms, a recursive predicate of arity $n$ requires enumerating $k^n$
  choices for $\vec{E}$).  A reasonable heuristic is to let $\Pi$ be the
  strongest pure formula implied by $S$, and enumerate only those combinations
  of $Z$ and $\vec{E}$ such that there is some $\Pi' :\Sigma' \in \mathit{cases}(Z(\vec{R},\vec{x}))$ such that
  $\ul{\Pi'}[\vec{E}/\vec{x}] \land \Pi \land x \neq y$ is unsatisfiable. 
  For
  example, for \texttt{assume(x $\neq$ y)}, this heuristic means that we
  enumerate only $\tuple{x,y}$ and $\tuple{y,x}$ (i.e, we attempt to introduce
  a list segment from $x$ to $y$ and from $y$ to $x$).
\shortenpar

We conclude this section with a theorem stating the correctness of our spatial
interpolation procedure.
\begin{theorem} \label{thm:spint_sound}
  Let $S$ and $I'$ be \sep formulas and let \texttt{c} be a command such that
  $\post(\texttt{c},S) \vdash I'$.  Then $\interp{S}{\texttt{c}}{I'}$ is a spatial interpolant
  for $S$, \texttt{c}, and $I'$.
\end{theorem}

%% file: snint.tex
\section{Spatial Interpolation Modulo Theories} \label{sec:snint}

\begin{figure}[htp]
\centering
\figsep{Entailment rules}
\vspace{-.15in}
\begin{mathpar}
  
  \inferrule[Star]{
    \cjudge{\Pi \land \Phi : \Sigma_0 \vdash \Pi' \land \Phi' : \Sigma_0'}
           { \mathcal{C}_0} \\
    \cjudge{\Pi \land \Phi : \Sigma_1 \vdash \Pi' \land \Phi' : \Sigma_1'}
           { \mathcal{C}_1 }  
  }{
    \cjudge{\Pi \land \Phi : \Sigma_0 * \Sigma_1  \vdash \Pi' \land \Phi' : \Sigma_0' * \Sigma_1'}
           {\mathcal{C}_0;\mathcal{C}_1}
  }

  \inferrule[Points-to]{ \Pi \models \Pi' } {
    \cjudge{\Pi \land \Phi : E \mapsto [\vec{A},\vec{F}] \vdash \Pi' \land \Phi' : E \mapsto [\vec{A},\vec{F}]}
           { \Phi' \gets \Phi }
  }

\inferrule*[lab=Fold,right={\rm $P \in \mathit{cases}(Z(\vec{R},\vec{x}))$}]{
    \cjudge{\Pi : \Sigma \vdash \Pi' : \Sigma' * P[\vec{\tau}/\vec{R},\vec{E}/\vec{x}]}
           {\mathcal{C}}
  }{
    \cjudge{\Pi : \Sigma \vdash \Pi' : \Sigma' * Z(\vec{\tau},\vec{E}) }
           {\mathcal{C}}
  }

\inferrule*[lab=Unfold,right={\rm\begin{minipage}{3cm}$\{P_1,\dotsc,P_n\} =$\\$\mathit{cases}(Z(\vec{R},\vec{x}))$\end{minipage}}]{
    \cjudge{\Pi : \Sigma * P_1[\vec{\tau}/\vec{R},\vec{E}/\vec{x}] \vdash \Pi' : \Sigma' }
           {\mathcal{C}_1}\\
           \dotsi\\\\
    \cjudge{\Pi : \Sigma * P_n[\vec{\tau}/\vec{R},\vec{E}/\vec{x}] \vdash \Pi' : \Sigma'}
           {\mathcal{C}_n}
  }{
    \cjudge{\Pi : \Sigma * Z(\vec{\tau},\vec{E}) \vdash \Pi' : \Sigma'}
           {\mathcal{C}_1; \dotsc{;}\, \mathcal{C}_n}
  }

  \inferrule*[lab=Predicate,right={\rm\begin{minipage}{3cm}
        Where $\tau_i = \lda{\vec{a}_i. \Psi_i}$\\
        and $\tau_i' = \lda{\vec{a}_i. \Psi_i'}$
      \end{minipage}}]{
    \Pi \models \Pi' \\
  }{
    \Phi' \gets \Phi; \Psi_1' \gets \Psi_1 \land \Phi; \dotsc{;}\, \Psi_{|\vec{\tau}|}' \gets \Psi_{|\vec{\tau}|} \land \Phi
    \triangleright\\
    \Pi \land \Phi : Z(\vec{\tau},\vec{E}) \vdash \Pi' \land \Phi' : Z(\vec{\tau'},\vec{E})
  }
\end{mathpar}
\vspace{-.05in}
\figsep{Execution rules}
\vspace{-.15in}
\begin{mathpar}
  \inferrule[Data-Assume]{
    \cjudge{P \land \phi \vdash Q}
           {\mathcal{C}}
  }{
    \cjudge{\hoare{P}
                  {assume($\phi$)}
                  {Q}}
           {\mathcal{C}}
  }

  \inferrule[Free]{
    \cjudge{P \vdash \Pi \land \Phi : \Sigma * x \mapsto [\vec{A},\vec{E}]}
           {\mathcal{C}}
  }{
    \cjudge{\hoare{P}{free(x)}{\Pi \land \Phi : \Sigma}}
           {\mathcal{C}}
  }

  \inferrule[Sequence]{
    \cjudge{\hoare{P}
                  {$\pi_0$}
                  {\widehat{O}}}
           {\mathcal{C}_0}\\
    \cjudge{\hoare{\widehat{O}}
                  {$\pi_1$}
                  {Q}}
           {\mathcal{C}_1}
  }{
    \cjudge{\hoare{P}
                  {$\pi_0;\pi_1$}
                  {Q}}
           {\mathcal{C}_0;\mathcal{C}_1}
  }

  \inferrule[Data-Load]{
    \cjudge{P \vdash \qex{X}{\Pi \land \widehat{\Phi} : \widehat{\Sigma} * x \mapsto [\vec{A},\vec{E}]}}
           {\mathcal{C}_0}\\\\
    \cjudge{\qex{X,a'}{\Pi[a'/a] \land \widehat{\Phi}[a'/a] \land a = A_i[a'/a] \colon (\widehat{\Sigma} * x \mapsto [\vec{A},\vec{E}])[a'/a]} \vdash Q}
           {\mathcal{C}_1}         
  }{
    \cjudge{\hoare{P}
                  {a := x->D$_i$}
                  {Q}}
           {\mathcal{C}_0; \mathcal{C}_1 }
  }

  \inferrule[Data-Assign]{
    \cjudge{\qex{a'}{\Pi \land \Phi[a'/a] \land a = A[a'/a] : \Sigma[a'/a] \vdash Q}}
           {\mathcal{C}}
  }{
    \cjudge{\hoare{\Pi \land \Phi \colon \Sigma}{a := A}{Q}}
           { \mathcal{C} }
  }

  \inferrule[Data-Store]{
    \cjudge{P \vdash \qex{X}{\Pi \land \widehat{\Phi} : \widehat{\Sigma} * x \mapsto [\vec{A},\vec{E}]}}
           {\mathcal{C}_0}\\\\
    \cjudge{\qex{X,a'}{\Pi \land \widehat{\Phi} \land a' = A : \widehat{\Sigma} * x \mapsto [\vec{A}[a'/A_i],\vec{E}]} \vdash Q}
           {\mathcal{C}_1}
  }{
    \cjudge{\hoare{P}{x->D$_i$ := A}{Q}}
           {\mathcal{C}_0; \mathcal{C}_1}
  }

  \inferrule[Alloc]{
    \cjudge{\qex{x',\vec{a},\vec{x}}{\Pi[x'/x] \land \Phi : \Sigma[x'/x] * x \mapsto [\vec{a},\vec{x}]} \vdash Q}
           {\mathcal{C}}
  }{
    \cjudge{\hoare{\Pi \land \Phi \colon \Sigma}
                  {x := new($n$,$m$)}
                  {Q}}
           {\mathcal{C}}
\vspace{-19pt}
  }
\end{mathpar}
\caption{Constraint generation.} 
\label{fig:constraints}
\end{figure}

\newcolumntype{L}[1]{>{\raggedright\let\newline\\\arraybackslash\hspace{0pt}}m{#1}}
\newcolumntype{C}[1]{>{\centering\let\newline\\\arraybackslash\hspace{0pt}}m{#1}}
\newcolumntype{R}[1]{>{\raggedleft\let\newline\\\arraybackslash\hspace{0pt}}m{#1}}

\begin{figure}
  \setlength{\abovecaptionskip}{0pt}
  \setlength{\belowcaptionskip}{4pt}
   \small
   \scalebox{0.8}{
   \begin{minipage}[t]{5.5cm}
     \textbf{Refined memory safety proof $\zeta'$ }\\
     $\color{dgray} \{R_0(i):\true\}$\\
     \texttt{i = nondet(); x = null}\\
     $\color{dgray}  \{R_1(i):\ls(\lda{a. R_{\ls 1}(\nu,i)},x,\nil) * \true\}$\\
     \texttt{assume(i != 0); \ldots ; i-\!-; }\\
     $\color{dgray} \{R_2(i):\ls(\lda{a. R_{\ls 2}(\nu,i)},x,\nil) * \true\}$\\
     \texttt{assume(i == 0)}\\
     $\color{dgray}  \{R_3(i):\ls(\lda{a. R_{\ls 3}(\nu,i)},x,\nil) * \true\}$\\
     \texttt{assume(x != null)}\\
     $\color{dgray}  \{\qex{d',y}{R_4(i,d'): x \mapsto [d',y] * \true}\}$\\
   \end{minipage}}
   \scalebox{0.8}{
   \begin{minipage}[t]{6.7cm}
     \textbf{Constraint system $\mathcal{C}$}\\
     $R_0(i') \gets \textit{true}$\\
     $R_1(i') \gets R_0(i)$\\
     $R_2(i') \gets R_1(i) \land i \neq 0 \land i' = i + 1$\\
     $R_3(i) \gets R_2(i) \land i = 0$\\
     $R_4(i,d') \gets R_3(i) \land R_{\ls 3}(d',i)$\\
     $R_{\ls 2}(\nu,i') \gets R_1(i) \land R_{\ls 1}(\nu,i) \land i \neq 0 \land i' = i + 1$\\
     $R_{\ls 2}(\nu,i') \gets R_1(i) \land \nu = i \land i \neq 0 \land i' = i + 1$\\
     $R_{\ls 3}(\nu,i) \gets R_2(i) \land R_{\ls 2}(\nu,i) \land i = 0$\\
     $d' \geq 0 \gets R_4(i,d')$
   \end{minipage}}
   \scalebox{0.8}{
   \begin{minipage}[t]{2.5cm}
     \textbf{Solution $\sigma$}\\
     $R_0(i): \textit{true}$\\
     $R_1(i): \textit{true}$\\
     $R_2(i): \textit{true}$\\
     $R_3(i): \textit{true}$\\
     $R_4(i,d'): d' \geq 0$\\
     $R_{\ls 1}(\nu,i): \nu \geq i$\\
     $R_{\ls 2}(\nu,i): \nu \geq i$\\
     $R_{\ls 3}(\nu,i): \nu \geq 0$
   \end{minipage}}
   \caption{Example constraints. \label{fig:ex_refine}}
 \end{figure}

We now consider the problem of \emph{refining} (or \emph{strengthening}) a
given separation logic proof of memory safety with information about
(non-spatial) data.  This refinement procedure results in a proof of a
conclusion stronger than can be proved by reasoning about the heap alone.  In
view of our example from Fig.~\ref{fig:ex}, this section addresses how to
derive the third sequence (Spatial Interpolants Modulo Theories) from the
second (Spatial Interpolants).

The input to our spatial interpolation modulo theories procedure is a
path $\pi$, a separation logic (\sep) proof $\zeta$ of the triple
$\hoare{\textit{true}:\emp}{$\pi$}{\textit{true}:\true}$ (i.e., a memory
safety proof for $\pi$), and a postcondition $\phi$.  The goal is to
transform $\zeta$ into an \logic proof of the triple
$\hoare{\textit{true}:\emp}{$\pi$}{\phi:\true}$.  The high-level operation of
our procedure is as follows.  First, we traverse the memory safety proof
$\zeta$ and build (1) a corresponding \emph{refined} proof $\zeta'$ where
refinements may contain second-order variables, and (2) a constraint system
$\mathcal{C}$ which encodes logical dependencies between the second-order
variables.  We then attempt to find a solution to $\mathcal{C}$, which is an
assignment of data formulas to the second-order variables such that all
constraints are satisfied.  If we are successful, we use the solution to
instantiate the second-order variables in $\zeta'$, which yields a valid
\logic proof of the triple $\hoare{\textit{true}:\emp}{$\pi$}{\phi:\true}$.
\shortenpar

\paragraph{Horn Clauses} The constraint system
produced by our procedure is a recursion-free set of Horn clauses, which can
be solved efficiently using existing first-order interpolation
techniques (see~\cite{Rummer13} for a detailed survey).
Following \cite{Gupta2011}, we define a \emph{query} to be an application
$Q(\vec{a})$ of a second-order variable $Q$ to a vector of (data) variables,
and define an \emph{atom} to be either a data formula $\phi \in
\textsf{DFormula}$ or a query $Q(\vec{a})$.  A \emph{Horn clause} is of the
form
\iflong\[\else $\fi
h \gets b_1 \land \cdots \land b_N
\iflong\]\else$ \fi
where each of $h,b_1,\dotsc,b_N$ is an atom.  In our constraint generation rules,
it will be convenient to use a more general form
which can be
translated to Horn clauses: we will allow constraints of the form
\iflong\[\else $\fi
h_1 \land \cdots \land h_M \gets b_1 \land \cdots \land b_N
\iflong\]\else$ \fi
(shorthand for the set of Horn clauses $\{ h_i \gets b_1 \land \cdots
\land b_N \}_{1 \leq i \leq M}$)
and we will allow queries to be of the form $Q(\vec{A})$ (i.e., take arbitrary
data terms as arguments rather than variables).
If $\mathcal{C}$ and $\mathcal{C}'$ are sets of constraints, we will use
$\mathcal{C};\mathcal{C}'$ to denote their union.

A \emph{solution} to a system of Horn clauses $\mathcal{C}$ is a map $\sigma$
that assigns each second-order variable $Q$ of arity $k$ a $\textsf{DFormula}$
$Q^\sigma$ with free variables drawn from $\vec{\nu} =
\tuple{\nu_1,\dotsc,\nu_k}$ such that for each clause
\iflong\[\else $\fi
h \gets b_1 \land \cdots \land b_N
\iflong\]\else$ \fi
in $\mathcal{C}$ the implication
\iflong\[\else $\fi
\forall A. (h^\sigma \Leftarrow (\exists B. b_1^\sigma \land \cdots \land b_N^\sigma)) 
\iflong\]\else$ \fi
holds, where $A$ is the set of free variables in $h$ and $B$ is the set of
variables
free in some $b_i$ but not in $h$.  In the above, for any
data formula $\phi$, $\phi^\sigma$ is defined to be $\phi$, and for any query
$Q(\vec{a})$, $Q(\vec{a})^\sigma$ is defined to be
$Q^\sigma[a_1/\nu_1,\dotsc,a_k/\nu_k]$ (where $k$ is the arity of $Q$).
\shortenpar

\paragraph{Constraint Generation Calculus}
We will present our algorithm for spatial interpolation modulo theories as a
calculus whose inference rules mirror the ones of separation logic.  The
calculus makes use of the same syntax used in recursive predicate definitions
in Sec.~\ref{sec:prelims}.  We use $\tau$ to denote a \emph{refinement
  term} and $\Phi$ to denote a \emph{refined formula}.  The calculus has two
types of judgements.  An \emph{entailment judgement} is of the form
\[
  \cjudge{\qex{X}{\Pi \land \Phi : \Sigma} \vdash \qex{X'}{\Pi' \land \Phi' : \Sigma'}}{\mathcal{C}}
\]
where $\Pi,\Pi'$ are equational pure assertions over heap terms,
$\Sigma,\Sigma'$ are refined spatial assertions, $\Phi$, $\Phi'$ are
refined formulas, and $\mathcal{C}$ is a recursion-free set of Horn clauses.
Such an entailment judgement should be read as ``for any solution $\sigma$ to
the set of constraints $\mathcal{C}$, $\qex{X}{\Pi \land \Phi^\sigma : \Sigma^\sigma}$
entails $\qex{X'}{\Pi' \land \Phi'^\sigma : \Sigma'^\sigma}$,''
where $\Phi^\sigma$ is $\Phi$ with all second order variables
replaced by their data formula assignments in $\sigma$ (and similarly
for $\Sigma^\sigma$).

Similarly, an \emph{execution judgement} is of the form
\[
  \cjudge{\hoare{\qex{X}{\Pi \land \Phi : \Sigma}}{$\pi$}{\qex{X'}{\Pi' \land \Phi' : \Sigma'}}}{\mathcal{C}}
\]
where $\pi$ is a path and $X,X',\Pi,\Pi',\Phi,\Phi',\Sigma,\Sigma'$, and $\mathcal{C}$
are as above.  Such an execution judgement should be read as ``for any
solution $\sigma$ to the set of constraints $\mathcal{C}$,\shortenpar \[\hoare{\qex{X}{\Pi \land \Phi^\sigma
: \Sigma^\sigma}}{\ensuremath \pi}{\qex{X'}{\Pi' \land \Phi'^\sigma : \Sigma'^\sigma}}\] is a valid triple.''

Let $\pi$ be a path, let $\zeta$ be a separation logic proof of the triple
$\hoare{\textit{true}:\emp}{$\pi$}{\textit{true}:\true}$ (i.e., a memory
safety proof for $\pi$), and let $\phi \in \textsf{DFormula}$ be a
postcondition.  Given these inputs, our algorithm operates as follows.  We
use $\vec{v}$ to denote a vector of all data-typed program variables.
The triple is \emph{rewritten with refinements} by letting $R$ and $R'$ be
fresh second-order variables of arity $|\vec{v}|$ and conjoining $R(\vec{v})$
and $R'(\vec{v})$ to the pre and post.
By recursing on $\zeta$, at each step applying the appropriate rule from our
calculus in Fig.~\ref{fig:constraints}, we derive a judgement\shortenpar
\begin{mathpar}
  \text{\footnotesize{\(
  \inferrule{
    \zeta'
  }{
    \cjudge{\hoare{\textit{true} \land R(\vec{v}):\true}{$\pi$}{\textit{true} \land R'(\vec{v}):\true}}{\mathcal{C}}
  }
  \)}}
\end{mathpar}
and then compute a solution $\sigma$ to the constraint
system \[\mathcal{C};\hspace*{0.25cm} R(\vec{v}) \gets
\textit{true};\hspace*{0.25cm} \phi \gets R'(\vec{v})\] (if one exists).  The
algorithm then returns $\zeta'^\sigma$, the proof obtained by applying the
substitution $\sigma$ to $\zeta'$.

Intuitively, our algorithm operates by recursing on a separation logic proof,
introducing refinements into formulas on the way down, and building a system
of constraints on the way up.  Each inference rule in the calculus encodes
both the downwards and upwards step of this algorithm.  For example, consider
the {\sc Fold} rule of our calculus:
we will illustrate the intended reading of this rule with a concrete example.
Suppose that the input to the algorithm is a derivation of the following form:
\begin{mathpar}
  \text{\footnotesize{\(
  \inferrule*[right=Fold]{
    \inferrule*{
      \zeta_0
    }{
      x \mapsto [a,\nil] \vdash \qex{b,y}{x \mapsto [b,y] * \ls(y,\nil)}
    }
  }{
    Q(i) : x \mapsto [a,\nil] \vdash R(i) : \ls(\lda{a. S(x,a)}, x,\nil)
  }
  \)}}
\end{mathpar}
(i.e., a derivation where the last inference rule is an application of {\sc
  Fold}, and the conclusion has already been rewritten with
refinements).  We introduce refinements in the premise and recurse
on the following derivation:
\begin{mathpar}
  \text{\footnotesize{\(
  \inferrule*{
    \zeta_0
  }{
    Q(i) : x \mapsto [a,\nil] \vdash \\\qex{b,y}{R(i) \land S(i,b) : x \mapsto [b,y] * \ls(\lda{a. S(x,a)},y,\nil)}
  }
  \)}}
\end{mathpar}
The result of this recursive call is a refined derivation $\zeta_0'$ as well as
a constraint system $\mathcal{C}$.  We then return both (1) the refined
derivation obtained by catenating the conclusion of the {\sc Fold} rule onto
$\zeta_0'$ and (2) the constraint system $\mathcal{C}$.

A crucial point of our algorithm is hidden inside the hat notation in
Fig.~\ref{fig:constraints} (e.g, $\widehat{O}$ in {\sc Sequence}): this
notation is used to denote the introduction of fresh second-order variables.
For many of the inference rules (such as {\sc Fold}), the refinements which
appear in the premises follow fairly directly from the refinements which
appear in the conclusion.  However, in some rules entirely new formulas appear
in the premises which do not appear in the conclusion (e.g., in the {\sc
  Sequence} rule in Fig.~\ref{fig:constraints}, the intermediate assertion
$\widehat{O}$ is an arbitrary formula which has no obvious relationship to the
precondition $P$ or the postcondition $Q$).  We refine such formula $O$ by
introducing a fresh second-order variable for the pure assertion and for each
refinement term that appears in $O$.
The following offers a concrete example.

\begin{example}
Consider the trace $\pi$ in Fig.~\ref{fig:ex}.
Suppose that we are given a memory safety proof for $\pi$ which ends in an
application of the {\sc Sequence} rule:
\begin{mathpar}
  \text{\footnotesize{\(
  \inferrule*[right=Sequence]{
    \hoare{\textit{true} : \emp}{$\pi_0$}{\textit{true} : \ls(x,\nil)}\\
    \hoare{\textit{true} : \ls(x,\nil)}{$\pi_1$}{\qex{b,y}{\textit{true} : x \mapsto [b,y]}}
  }{
    \hoare{Q(i) : \emp}{$\pi_0;\pi_1$}{\qex{b,y}{R(i,b) : x \mapsto [b,y]}}
  }
  \)}}
\end{mathpar}
where $\pi$ is decomposed as $\pi_0;\pi_1$, $\pi_0$ is the path from 1 to 3,
and $\pi_1$ is the path from 3 to 4.  Let $O = \textit{true} : \ls(x,\nil)$
denote the intermediate assertion which appears in this proof. To derive
$\widehat{O}$, we introduce two fresh second order variables, $S$ (with arity
1) and $T$ (with arity 2), and define $\widehat{O} = {S(i) :
    \ls(\lda{a. T(i,a)},x,\nil)}$.  The resulting inference is as follows:
\begin{mathpar}
  \text{\footnotesize{\(
  \inferrule*{
    \hoare{Q(i) : \emp}{$\pi_0$}{S(i) : \ls(\lda{a.T(i,a)},x,\nil)}\\
    \hoare{S(i) : \ls(\lda{a.T(i,a)},x,\nil)}{$\pi_1$}{\qex{b,y}{R(i,b) : x \mapsto [b,y]}}
  }{
    \qquad
    \hoare{Q(i) : \emp}{$\pi_0;\pi_1$}{\qex{b,y}{R(i,b) : x \mapsto [b,y]}}
    \qquad\lrcorner
  }
  \)}}
\end{mathpar}
\end{example}

The following example provides a simple demonstration of our
constraint generation procedure:

\begin{example}
Recall the example in Fig.~\ref{fig:ex} of Sec.~\ref{sec:ex}.
The row of spatial interpolants in Fig.~\ref{fig:ex} is a memory safety proof $\zeta$
of the program path.
Fig.~\ref{fig:ex_refine} shows the refined proof $\zeta'$,
which is the proof $\zeta$ with second-order variables
that act as placeholders for data formulas.
\textbf{\emph{For the sake of illustration, we have simplified the constraints
by skipping a number of intermediate annotations in the Hoare-style proof.}}

The constraint system $\mathcal{C}$ specifies the logical dependencies
between the introduced second-order variables in $\zeta'$.
For instance, the relation between $R_{2}$ and $R_{3}$ is
specified by the Horn clause  
     $R_{3}(i) \gets R_{2}(i) \land i = 0$,
which takes into account the constraint imposed by 
$\texttt{assume (i == 0)}$ in the path.
The Horn clause 
     $d' \geq 0 \gets R_4(i,d')$
specifies the postcondition defined by the assertion
$\texttt{assert(x->D >= 0)}$,
which states that the value of the data field
of the node $x$ should be $\geq 0$.
\shortenpar

Replacing second-order variables in $\zeta'$ with their respective
solutions in $\sigma$ produces
a proof that the assertion
at the end of the path holds (last row of Fig.~\ref{fig:ex}).
\eoe\end{example}

\paragraph{Soundness and Completeness}
The key result regarding the constraint systems produced by these judgements
is that any solution to the constraints yields a valid refined proof.  The
formalization of the result is the following theorem.
\begin{theorem}[Soundness] \label{thm:refine_sound}
  Suppose that $\pi$ is a path, $\zeta$ is a derivation of the judgement
  $\cjudge{\hoare{P}{$\pi$}{Q}}{\mathcal{C}}$, and that
  $\sigma$ is a solution to $\mathcal{C}$.  Then $\zeta^\sigma$, the proof
  obtained by applying the substitution $\sigma$ to $\zeta$, is a (refined)
  separation logic proof of $\hoare{P^\sigma}{$\pi$}{Q^\sigma}$.
\end{theorem}

Another crucial result for our counterexample generation strategy is
a kind of completeness theorem, which effectively states that the strongest memory safety proof always admits a refinement.
\begin{theorem}[Completeness] \label{thm:refine_complete}
  Suppose that $\pi$ is a memory-feasible path and $\zeta$ is a derivation of the judgement
  $\cjudge{\hoare{R_0(\vec{v}):\emp}{$\pi$}{R_1(\vec{v}) : \true}}{\mathcal{C}}$ obtained by symbolic execution.  If $\phi$ is a data formula such that 
  $\hoare{\textit{true}:\emp}{$\pi$}{\phi : \true}$
  holds, then there is a solution $\sigma$ to $\mathcal{C}$ such that
  $R_1^\sigma(\vec{v}) \Rightarrow \phi$.
\end{theorem}

%% file: bounded_abduction.tex
\section{Bounded Abduction} \label{sec:abduction}
\newcommand{\coloured}[2]{%
  \ifthenelse{\equal{#2}{r}}{%
    {\color{BrickRed}{\ensuremath[#1]^{\rm #2}}}
  }{\ifthenelse{\equal{#2}{b}}{%
    {\color{RoyalBlue}\ensuremath[#1]^{\rm #2}}
    }{%
    \ensuremath[#1]^{#2}
    }
  }
}
\newcommand{\hiding}[2]{\ensuremath\langle#1 \unlhd #2\rangle}

\begin{figure*}[t]
\setlength{\abovecaptionskip}{-4pt}
\setlength{\belowcaptionskip}{4pt}
\centering
\begin{mathpar}
 \scriptsize
  \inferrule[Empty]{\Pi \models \Pi'}{
    \Pi : \coloured{\emp}{c} \vdash \Pi' : \hiding{\coloured{\emp}{c}}{\emp}
  }

  \inferrule[Star]{
    \Pi : \Sigma_0 \vdash \Pi' : \Sigma_0' \\
    \Pi : \Sigma_1 \vdash \Pi' : \Sigma_1'
  }{
    \Pi : \Sigma_0 * \Sigma_1  \vdash \Pi' : \Sigma_0' * \Sigma_1'
  }

   \inferrule[Points-to]{\Pi \models \Pi'}{
     \Pi : \coloured{E \mapsto [a,F]}{c}
     \vdash
     \Pi' : \hiding{\coloured{E \mapsto [a, F]}{c}}{E \mapsto [a, F]}
  }

  \inferrule[True]{\Pi \models \Pi'}{
    \Pi : \Sigma
    \vdash
    \Pi' : \hiding{\coloured{\true}{c}}{\true}
  }

  \inferrule[Substitution]{
    \Pi[E/x] : \Sigma[E/x] \vdash \Pi'[E/x] : \Sigma'[E/x] \\
    \Pi \models x = E
  }{
    \Pi : \Sigma \vdash \Pi' : \Sigma'
  }

 \inferrule[$\exists$-right]{
   P \vdash Q[\AE/x]
 }{
   P \vdash \qex{x}{Q}
 }
%
%
%
%
%
%
\end{mathpar}
\caption{Coloured strengthening.  All primed variables are chosen fresh.}
\label{fig:coloured_strengthening}
\end{figure*}

In this section, we discuss our algorithm for bounded abduction.  Given a
bounded abduction problem
\[L \vdash \qex{X}{M * [\ ]} \vdash R\]
we would like to find a formula $A$
such that $L \vdash \qex{X}{M * A} \vdash R$.
Our algorithm is sound but not complete: it is possible that there exists a
solution to the bounded abduction problem, but our procedure cannot find it.
In fact, there is in general no complete procedure for bounded abduction, as a
consequence of the fact that we do not pre-suppose that our proof system for
entailment is complete, or even that entailment is decidable.

\paragraph{High level description} Our algorithm proceeds in three steps:
\begin{compactenum}
\item Find a \emph{colouring} of $L$.  This is an assignment of a colour, either \emph{red} or \emph{blue},
  to each heaplet appearing in $L$.
  Intuitively, red heaplets are used to satisfy $M$, and blue heaplets are
  left over.  This colouring can be computed by recursion on a proof of $L
  \vdash \qex{X}{M * \true}$.
\item Find a \emph{coloured strengthening} $\Pi : \coloured{M'}{r} *
  \coloured{A}{b}$ of $R$. 
  (We use the notation $\coloured{\Sigma}{r}$ or $\coloured{\Sigma}{b}$ 
  to denote a spatial formula $\Sigma$ of red or blue colour, respectively.)
  Intuitively, this is a formula that (1) entails $R$
  and  (2) is coloured in such a way that the red heaplets correspond to the
  red heaplets of $L$, and the blue heaplets correspond to the blue heaplets
  of $L$.  This coloured strengthening can be computed by recursion on a proof
  of $L \vdash R$ using the colouring of $L$ computed in step 1.
\item Check
  $\Pi' : M * A \models R$, where $\Pi'$ is the
  strongest pure formula implied by $L$.  This step is necessary because $M$
  may be weaker than $M'$.  If the entailment check fails, then our algorithm
  fails to compute a solution to the bounded abduction problem.  If the
  entailment check succeeds, then $\Pi'' : A$ is a solution, where $\Pi''$ is
  the set of all equalities and disequalities in $\Pi'$ which were actually
  used in the proof of the entailment $\Pi' : M * A \models R$
  (roughly, all those equalities and disequalities which appear in the leaves of
  the proof tree, plus the equalities that were used in some instance of the
  {\sc Substitution} rule).
\shortenpar
\end{compactenum}

First, we give an example to illustrate these high-level steps:
\begin{example}
  Suppose we want to solve the following bounded abduction problem:
  \[ L \vdash \ls(x,y) * [\ ] \vdash R \]
  where $L = x \mapsto [a,y] * y \mapsto [b,\nil]$ and
    $R = \qex{z}{x\mapsto [a,z] * \ls(y,\nil)}$.
  Our algorithm operates as follows:
\begin{compactenum}
\item Colour $L$: $\coloured{x \mapsto [a,y]}{r} * \coloured{y \mapsto [b,\nil]}{b}$
\item Colour $R$: $\qex{z}{\coloured{x \mapsto [a,z]}{r} * \coloured{\ls(y,\nil)}{b}}$
\item Prove the entailment
  \[ x \neq \nil \land y \neq \nil \land x \neq y : \ls(x,y) * \ls(y,\nil) \models R\]
  This proof succeeds, and uses the pure assertion $x \neq y$.
\end{compactenum}
Our algorithm computes $x \neq y : \ls(y,\nil)$ as the solution to the bounded
abduction problem.
\eoe\end{example}

We now elaborate our bounded abduction algorithm.  We assume that
$L$ is quantifier free (without loss of generality, since quantified variables
can be Skolemized) and \emph{saturated} in the sense that for any pure
formula $\Pi'$, if $L \vdash \Pi'$, where $L = \Pi : \Sigma$, then $\Pi \vdash
\Pi'$.

\paragraph{Step 1} The first step of the algorithm is straightforward.  
If we suppose that there exists a
solution, $A$, to the bounded abduction problem, then by definition we must that have
$L \models \qex{X}{M * A}$.  Since $\qex{X}{M * A} \models \qex{X}{M *
  \true}$, we must also have $L \models \qex{X}{M * \true}$.  We begin step 1
by computing a proof of $L \vdash \qex{X}{M * \true}$.  If we fail, then we abort the procedure and report that we cannot find a solution to the abduction problem.
If we succeed, then we can
colour the heaplets of $L$ as follows: for each heaplet $E \mapsto
[\vec{A},\vec{F}]$ in $L$, either $E \mapsto [\vec{A},\vec{F}]$ was used in an
application of the {\sc Points-to} axiom in the proof of $L \vdash \qex{X}{M *
  \true}$ or not.  If yes, we colour $E \mapsto [\vec{A},\vec{F}]$ red;
otherwise, we colour it blue.  We denote a heaplet $H$ coloured by a
colour $c$ by $\coloured{H}{c}$.
\shortenpar

\paragraph{Step 2} The second step
is to find a coloured strengthening of $R$.  Again, supposing that there
is some solution $A$ to the bounded abduction problem, we must have $L \models
\qex{X}{M * A} \models R$, and therefore $L \models R$.  We begin step 2 by
computing a proof of $L \vdash R$.  If we fail, then we abort.  If we succeed, then we define a coloured
strengthening of $R$ by recursion on the proof of $L \vdash R$.  Intuitively,
this algorithm operates by inducing a colouring on points-to predicates in the
leaves of the proof tree from the colouring of $L$ (via the {\sc Points-to}
rule in Fig.~\ref{fig:coloured_strengthening}) and then only folding recursive predicates 
when all the folded heaplets have the same colour.
\shortenpar

More formally, for each formula $P$ appearing as the consequent of
 some sequent in a proof tree, our algorithm produces
 a mapping from heaplets in $P$ to coloured spatial formulas.  The mapping is represented using the
notation $\hiding{\Sigma}{H}$, which denotes that the heaplet $H$ is mapped to
the coloured spatial formula $\Sigma$.  For each recursive predicate $Z$ and each
$\qex{X}{\Pi : H_1 * \dotsi * H_n} \in \mathit{cases}(Z(\vec{R},\vec{x}))$, we define two versions
of the fold rule, corresponding to when $H_1,\ldots,H_n$ are coloured
homogeneously ({\sc Fold1}) and heterogeneously ({\sc Fold2}):
\begin{mathpar}
\scriptsize
  \inferrule[Fold1]{
    (\Pi : \Sigma \vdash \Pi' : \Sigma' * \hiding{\coloured{H_1}{c}}{H_1} * \dotsi * \hiding{\coloured{H_n}{c}}{H_n})[\vec{E}/\vec{x}]
  }{
    \Pi : \Sigma \vdash \Pi' : \Sigma' * \hiding{\coloured{Z(\vec{E})}{c}}{Z(\vec{E})}
  }

  \inferrule[Fold2]{
    (\Pi : \Sigma \vdash \Pi' : \Sigma' * \hiding{\Sigma_1'}{H_1} * \dotsi * \hiding{\Sigma_n'}{H_n})[\vec{E}/\vec{x}]
  }{
    \Pi : \Sigma \vdash \Pi' : \Sigma' * \hiding{\Sigma_1'* \dotsi * \Sigma_n'}{Z(\vec{E})}
  }
\end{mathpar}

The remaining rules for our algorithm are presented formally in
Fig.~\ref{fig:coloured_strengthening}.\footnote{Note that some of the
  inference rules are missing.  This is because these rules are inapplicable
  (in the case of {\sc Unfold} and {\sc Inconsistent}) or unnecessary (in the
  case of {\sc \nil-not-Lval} and {\sc*-Partial}), given our assumptions on
  the antecedent.}
To illustrate how this algorithm works, consider the {\sc Fold1} and {\sc
  Fold2} rules.  If a given (sub-)proof finishes with an instance of {\sc
  Fold} that
folds $H_1 *\cdots* H_n$ into $Z(\vec{E})$, we begin by colouring the
sub-proof of
\[\Pi : \Sigma \vdash \Pi' : \Sigma' * H_1 * \dotsi * H_n \]
This colouring process produces a coloured heaplet $\Sigma_i$ for each $H_i$.
If there is some colour $c$ such that each $\Sigma_i'$ is $\coloured{H_i}{c}$,
then we apply {\sc Fold1} and $Z(\vec{E})$ gets mapped to
$\coloured{Z(\vec{E})}{c}$.  Otherwise (if there is some $i$ such that
$\Sigma_i$ is not $H_i$ or there is some $i,j$ such that $\Sigma_i$ and
$\Sigma_j$ have different colours), we apply {\sc Fold2}, and map $Z(\vec{E})$
to $\Sigma_1* \dotsi *\Sigma_n$.

After colouring a proof, we define $A$ to be the blue part of $R$.  That is,
if the colouring process ends with a judgement of\\
$\Pi : \coloured{\Sigma_1}{r} * \coloured{\Sigma_2}{b} \vdash
\Pi' : \hiding{\coloured{\Sigma_{11}'}{r}*\coloured{\Sigma_{12}}{b}}{H_1} * \cdots * \hiding{\coloured{\Sigma_{n1}'}{r}*\coloured{\Sigma_{n2}}{b}}{H_n}$\\
(where for any coloured spatial formula $\Sigma$, its partition into red and
  blue heaplets is denoted by
  $\coloured{\Sigma_1}{r}*\coloured{\Sigma_2}{b}$), we define $A$ to be $\Pi'
  : \Sigma_{12} * \dotsi * \Sigma_{n2}$.  This choice is justified by the
  following lemma:

\begin{lemma}
  Suppose that\\
  $\Pi : \coloured{\Sigma_1}{r} * \coloured{\Sigma_2}{b} \vdash
  \Pi' : \hiding{\coloured{\Sigma_{11}'}{r}*\coloured{\Sigma_{12}}{b}}{H_1} * \cdots * \hiding{\coloured{\Sigma_{n1}'}{r}*\coloured{\Sigma_{n2}}{b}}{H_n}$\\
  is derivable using the rules of Fig.~\ref{fig:coloured_strengthening}, and
  that the antecedent is saturated.
  Then the following hold:
  \begin{compactitem}
  \item $\Pi' : \Sigma_{11}*\Sigma_{12}*\dotsi*\Sigma_{n2} \models \Pi' : H_1 * \dotsi * H_n$;
  \item $\Pi : \Sigma_1 \models \Pi' : \Sigma_{11} * \dotsi * \Sigma_{n1}$; and
  \item $\Pi : \Sigma_2 \models \Pi' : \Sigma_{12} * \dotsi * \Sigma_{n2}$.
  \end{compactitem}
\end{lemma}

\paragraph{Step 3} 
The third step of our algorithm is to check the entailment $\Pi : M * A \models R$.
To illustrate why this is necessary, consider the following example:
\begin{example}
Suppose we want to solve the following bounded abduction problem:
\[x \neq y : x \mapsto [a,y] \vdash \ls(x,y)*[\ ] \vdash x \mapsto [a,y]\ .\]
In Step 1, we compute the colouring $x \neq y : \coloured{x \mapsto
  [a,y]}{r}*\coloured{\emp}{b}$ of the left hand side.  In step 2, we compute
the colouring $\coloured{x \mapsto [a,y]}{r}*\coloured{\emp}{b}$ of the
right hand side.  However, $\emp$ is not a solution to the bounded abduction
problem.  In fact, there is no solution to the bounded abduction problem.
Intuitively, this is because $M$ is too weak to entail the red part of the
right hand side.
\eoe\end{example}

%% file: eval.tex
\section{Implementation and Evaluation} \label{sec:impl}

Our primary goal is to study the feasibility of our
proposed algorithm. To that end, we implemented
an instantiation of our generic algorithm 
with the linked list recursive predicate $\ls$ (as defined in Sec.~\ref{sec:prelims})
and refinements in the theory of linear arithmetic (QF\_LRA).
The following describes our implementation and evaluation of \alg in detail.

\paragraph{Implementation}
We implemented \alg in the T2 safety and termination verifier~\cite{t2}.
Specifically, we extended T2's front-end to handle heap-manipulating programs,
and used its safety checking component (which implements 
McMillan's \impact algorithm) as a basis for our implementation of \alg.
To enable reasoning in separation logic, 
we implemented an entailment checker for \logic
along with a bounded abduction procedure.

We implemented a constraint-based solver
using the linear rational arithmetic
interpolation techniques of Rybalchenko and Stokkermans~\cite{Rybalchenko07}
to solve the non-recursive Horn clauses
generated by \alg.
Although many off-the-shelf tools for interpolation exist (e.g.,~\cite{McMillan2011})
we implemented
our own solver for experimentation and evaluation purposes to
 allow us more flexibility in controlling
the forms of interpolants we are looking for.
We expect that \alg would perform even better using
these highly tuned interpolation engines.
\shortenpar

Our main goal is to evaluate the feasibility
of our proposed extension of interpolation-based
verification to heap and data reasoning, and not necessarily
to demonstrate performance improvements against other tools.
Nonetheless, we note that there are two tools that
target similar programs: (1) \textsc{Thor}~\cite{Magill2010},
which computes a memory safety proof and uses off-the-shelf
numerical verifiers to strengthen it, and (2) \textsc{Xisa}~\cite{Chang2008},
which combines shape and data abstract domains in an abstract interpretation framework.
\textsc{Thor} cannot compute arbitrary refinements of recursive predicates
(like the ones demonstrated here and required in our benchmarks)
unless they are manually supplied with the required theory predicates.
Instantiated with the right abstract data domains, \textsc{Xisa}
 can in principle handle most programs we target in our evaluation.
(\textsc{Xisa} was unavailable for comparison~\cite{chang}.)
Sec.~\ref{sec:rel} provides  a detailed comparison
with related work.

\paragraph{Benchmarks}
To evaluate \alg, we used a number of
linked list benchmarks that require heap and data reasoning.
First, we used a number of simple benchmarks:
\texttt{listdata} is similar to Fig.~\ref{code:ex},
where a linked list is constructed and its data elements are later checked;
\texttt{twolists} requires an invariant comparing data elements of two lists
(all elements in list $A$ are greater than those in list $B$);
\texttt{ptloop} tests our spatial interpolation technique, where
the head of the list must not be folded in order to ensure its data element
is accessible;
and \texttt{refCount} is a reference counting program, where our goal
is to prove memory safety (no double free).
For our second set of benchmarks, we used a cut-down version of 
BinChunker (\url{http://he.fi/bchunk/}),
a Linux utility for converting between different audio CD formats.
BinChunker maintains linked lists and uses their data elements for traversing
an array. Our property of interest is thus ensuring that all array accesses are within bounds.
To test our approach, we used a number of modifications of BinChunker,
\texttt{bchunk\_a} to \texttt{bchunk\_f}, where \texttt{a} is the simplest
benchmark and \texttt{f} is the most complex one.

\paragraph{Heuristics}
We employed a number of heuristics to improve
our implementation.
First, given a program path to prove correct, 
we attempt to find a similar
proof to previously proven paths that 
traverse the same control flow locations.
This is similar to the \emph{forced covering} heuristic
of~\cite{McMillan2006} to force path interpolants
to generalize to inductive invariants.
Second, our Horn clause solver uses Farkas' lemma
to compute linear arithmetic interpolants.
We found that minimizing the number of non-zero \emph{Farkas coefficients}
results in more generalizable refinements. A similar heuristic
is employed by~\cite{Albarghouthi2013}.

\begin{table}[t]
\centering
\scalebox{0.8}{
\begin{tabular}{c|c|c|c|c}
Benchmark & \#\ppath &  Time (s)  & $\theory$ Time & Sp. Time \\ 
\hline
\hline
\texttt{listdata} & 5 & 1.37 & 0.45  &  0.2  \\ 
\texttt{twolists} & 5 & 3.12 & 2.06 & 0.27 \\
\texttt{ptloop} & 3 & 1.03 & 0.28 & 0.15\\
\texttt{refCount} & 14 & 1.6 & 0.59 & 0.14\\
\hline
\texttt{bchunk\_a} & 6 & 1.56 & 0.51 & 0.25\\
\texttt{bchunk\_b} & 18 & 4.78  & 1.7  & 0.2 \\
\texttt{bchunk\_c} & 69 & 31.6 & 14.3 & 0.26\\
\texttt{bchunk\_d} & 23 & 9.3 & 4.42 & 0.27 \\
\texttt{bchunk\_e} & 52 & 30.1 & 12.2 & 0.25 \\
\texttt{bchunk\_f} & 57 & 22.4 & 12.0 & 0.25 \\
\end{tabular}}
\caption{Results of running \alg on our benchmark set. }
\label{tbl:res}
\end{table}
\paragraph{Results}
Table~\ref{tbl:res} shows the results of 
running \alg on our benchmark suite.
Each row shows the number of calls to
$\ppath$ (number of paths proved),
the total time taken by \alg in seconds,
the time taken to generate
Horn clauses and compute theory interpolants ($\theory$ Time),
and the time taken to compute spatial interpolants (Sp. Time).
\alg proves all benchmarks correct w.r.t. their respective properties.
As expected, on simpler examples, the number of paths sampled
by \alg is relatively small (3 to 14).
In the \texttt{bchunk\_*} examples, \alg examines
up to 69 paths (\texttt{bchunk\_c}).
It is important to note that, in all benchmarks,
almost half of the total time is spent in
theory interpolation. We expect this can be drastically cut
 with the use of a more efficient interpolation engine.
The time taken by spatial interpolation is very small
in comparison, and becomes negligible in larger examples. 
The rest of the time is spent in checking entailment
of \logic formulas and other miscellaneous operations.

Our results highlight the utility of our proposed
approach. Using our prototype implementation of \alg,
we were able to verify a set of realistic
programs that require non-trivial combinations of 
heap and data reasoning.
We expect the performance of our prototype implementation of \alg can greatly
improve with the help of
state-of-the-art Horn clause solvers,
and more efficient entailment checkers for separation logic.

%% file: related.tex
\section{Related Work} \label{sec:rel}

\paragraph{Abstraction Refinement for the Heap}
To the best of our knowledge, the work of Botincan et al.~\cite{Botincan13} is the
only separation logic shape analysis that employs
a form of abstraction refinement. It
starts with a family of separation logic domains of increasing
precision, and uses
spurious counterexample traces (reported by forward fixed-point computation)
to pick a more precise domain
to restart the analysis and (possibly) eliminate the counterexample.
Limitations of this technique include:
(1) The precision of the analysis is contingent on the set of
abstract domains it is started with.
(2)
The refinement strategy (in contrast to \alg) does not guarantee progress (it may explore the same path repeatedly), and may report false positives.
On the other hand, given a program path, \alg is guaranteed to find a 
proof for the path or correctly declare it an unsafe execution.
(3) Finally, it is unclear whether refinement with a powerful
theory like linear arithmetic can be encoded in such a framework,
e.g., as a set of domains with increasingly more arithmetic predicates.

Podelski and Wies~\cite{Podelski10}
propose an abstraction refinement algorithm
for a shape-analysis domain with a logic-based
view of three-valued shape analysis (specifically,
first-order logic plus transitive closure).
Spurious counterexamples are used to either refine
the set of predicates used in the analysis,
or refine an imprecise abstract transformer.
The approach is used to verify specifications 
given by the user as first-order
logic formulas. A limitation of the approach
is that refinement is syntactic, and if an important
recursive predicate (e.g., there is a list from $x$ to $\nil$)
is not explicitly supplied in the specification, it cannot be inferred automatically.
Furthermore, abstract post computation can be expensive,
as the abstract domain uses quantified predicates.
Additionally, the analysis assumes a memory safe program to start,
whereas, in \alg, we construct a memory safety proof as part
of the invariant, enabling us to detect unsafe memory operations
that lead to undefined program behavior.

Beyer et al.~\cite{Beyer06} propose
using shape analysis information on demand 
to augment numerical predicate abstraction.
They use shape analysis
as a backup analysis when failing to prove
a given path safe without tracking the heap,
and incrementally refines TVLA's~\cite{Bogudlov07} three-valued shape analysis~\cite{Sagiv99} to track more heap information as required.
As with~\cite{Podelski10},~\cite{Beyer06} 
makes an \emph{a priori} assumption of memory safety and
requires an expensive abstract post operator.

Finally, Manevich et al.~\cite{Manevich06} give a theoretical treatment of counterexample-driven
refinement in power set (e.g., shape) abstract domains.

\paragraph{Combined Shape and Data Analyses}
The work of Magill et al.~\cite{Magill2010} infers shape and numerical
invariants, and is the most closely related to ours.  First, a separation logic
analysis is used to construct a memory safety proof of the whole program.  
This proof is then
\emph{instrumented} by adding additional user-defined integer parameters to
the recursive predicates appearing in the proof (with corresponding
user-defined interpretations).  
A numerical program is generated from this instrumented proof and checked using an off-the-shelf
verification tool, which need not reason about the heap.  
Our technique and \cite{Magill2010}'s are similar in that we both decorate
separation logic proofs with additional information: in \cite{Magill2010}, the
extra information is instrumentation variables; in this paper, the extra
information is refinement predicates.  Neither of these techniques properly
subsumes the other, and we believe that they may be profitably combined.  An
important difference is that we
synthesize data refinements automatically from program paths, whereas
\cite{Magill2010} uses a fixed (though user-definable) abstraction.

A number of papers have proposed
abstract domains for shape and data invariants.
Chang and Rival~\cite{Chang2008}
propose a separation logic--based abstract domain that is parameterized
by programmer-supplied \emph{invariant checkers} (recursive predicates)
and a data domain for reasoning about contents of these structures.
McCloskey et al.~\cite{DBLP:conf/sas/McCloskeyRS10} also proposed a
combination of heap and numeric abstract domains, this time using 3-valued
structures for the heap.
While the approaches to combining shape and data information are significantly different,
an advantage of our method is that it does not lose precision
due to limitations in the abstract domain, widening, and join.

Bouajjani et al.~\cite{Bouajjani10,Bouajjani12} propose
an abstract domain for list manipulating programs
that is parameterized by a data domain.
They show that by varying the data domain, one
can infer invariants about list sizes, sum of elements, etc.
Quantified data automata (QDA)~\cite{Garg13} have
been proposed as an abstract domain for representing
list invariants where the data in a list is described
by a regular language.
In~\cite{Garg13b}, invariants over QDA have been synthesized using
language learning techniques from concrete program executions.
Expressive logics have also been proposed for
reasoning about heap and data~\cite{Qiu13}, but have thus far 
been only used for invariant checking, not invariant synthesis.
A number of decision procedures for combinations of the 
singly-linked-list fragment of separation logic
with SMT theories have recently been proposed~\cite{Piskac13,Navarro13}.

\paragraph{Path-based Verification}
A number of works proposed path-based algorithms
for verification.
Our work builds on McMillan's \impact technique~\cite{McMillan2006}
and extends it to heap/data reasoning.
Earlier work~\cite{Henzinger04} used interpolants
to compute predicates from spurious paths in a CEGAR loop.
Beyer et al.~\cite{Beyer07} proposed \emph{path invariants},
where infeasible paths induce program slices
that are proved correct, and from which predicates are mined for full program verification.
Heizmann et al.~\cite{Heizmann10} presented a technique that uses
interpolants to compute path proofs
and generalize a path into a visibly push-down language of correct paths.
In comparison with \alg, 
all of these techniques are restricted to first-order invariants.
\shortenpar

Our work is similar to that of Itzhaky et al.~\cite{DBLP:conf/cav/ItzhakyBRST14}, in the sense that we both generalize from bounded unrollings of the program to compute ingredients of a proof.
However, they compute proofs in a fragment of first-order logic that can only express linked lists and has not yet been extended to combined heap and data properties.

%% file: alg.tex
\section{From Proofs of Paths to Proofs of Programs} \label{sec:alg}

In this section, we describe how \alg constructs a proof
of correctness (i.e., unreachability of the error location $v_{\rm e}$) of a program $\prog = \tuple{V,E,v_{\rm i},v_{\rm e}}$ from proofs of individual paths.
We note that our algorithm is an extension of \impact~\cite{McMillan2006}
to \logic proofs; we refer the reader to~\cite{McMillan2006}
for optimizations and heuristics.
The main difference is the procedure \ppath,
which constructs an \logic proof for a given program path.

\begin{figure}[t]
\centering

    \begin{algorithmic}[1]
    \Function{SplInter}{$\prog$}
        \State $S \gets \emptyset$
    \Loop \label{l:loop}
        \State $\pi \gets \cproof(S)$
        \If{$\pi$ is empty} // \emph{$S$ is a proof}
            \State \Return found proof $S$
        \Else ~~ // \emph{$\pi = v_{\rm i},\ldots,v_{\rm e}$ in $\prog$ does not appear in $S$}
            \State $\kappa \gets \ppath(\pi)$
            \If{$\kappa$ is empty} ~~ // \emph{No proof computed for $\pi$} 
                \State \Return found erroneous execution $\pi$
            \EndIf
            \State $S' \gets \emptyset$
            \For {each $\kappa' \in S$}
                \State $(\kappa,\kappa') \gets \conj(\kappa,\kappa')$
                \State $S' \gets S' \cup \{ \kappa' \}$
            \EndFor
            \State $S \gets S' \cup \{\kappa\}$
        \EndIf
    \EndLoop
    \EndFunction
    \end{algorithmic}
\caption{Main Algorithm of \alg.}
\label{alg:main}
\end{figure}

\paragraph{The Main Algorithm}
Figure~\ref{alg:main} shows the main algorithm of \alg.
Elements of the set $S$ are program paths
from $v_{\rm i}$ to $v_{\rm e}$, annotated with $\logic$ formulas.
For example, $$(a_1,v_1),(a_2,v_2),\ldots,(a_n,v_n)$$
is an annotated path where (1) $\{a_j\}_j$ are $\logic$ formulas;
(2) $v_1 = v_{\rm i}$ and $v_n = v_{\rm e}$;
(3) for $j \in [1,n-1]$, $(v_j,v_{j+1}) \in E$;
(4) for each edge $e = (v_j,v_{j+1})$, 
$\{a_j\}\;e^{\rm c}\;\{a_{j+1}\}$ is valid;
and (5) $a_n$ is $\false$
(since we are interested in proving unreachability of $v_{\rm e}$).

\alg uses the procedure $\cproof$ to check
if the set of annotated paths $S$ represents a
proof of correctness of the whole program.
If not, $\cproof$ samples a new program path ($\pi$)
that does not appear in $S$.
Using the procedure $\ppath$, it tries
to construct an annotation/proof (using spatial($\theory$) interpolants)
of $\pi$.
If no proof is constructed, \alg concludes
that $\pi$ is a feasible execution to $v_{\rm e}$ (or it performs
unsafe memory operations).
Otherwise, it uses the procedure $\conj$ to
strengthen the proofs of all program paths in $S$,
adds the annotated path to $S$, and restarts
the loop.

Note that the annotated paths in $S$ represent an
\emph{Abstract Reachability Tree} (ART), as used in~\cite{McMillan2006}
as well as other software model checking algorithms.
The tree in this case is rooted at $v_{\rm i}$, and branching
represents when two paths diverge.

We will now describe \alg's components in more detail.

\paragraph{\cproof: Checking Proofs}
Given a set of annotated paths $S$, 
we use the procedure $\cproof(S)$ to check if the
annotations represent a proof of the whole program.
Specifically,
for each $v \in V$, we use $I(v)$ to denote
the formula $\bigvee\{a_j \mid (a_j,v) \in \kappa, \kappa \in S\}$.
In other words, for each location $v$ in $\prog$,
we \emph{hypothesize} an inductive invariant $I(v)$ from our current annotations
of paths passing through $v$.
We can then check if our hypotheses indeed form
an inductive invariant of the program. If so, 
the program is safe:
since $I(v_{\rm e})$ is always $\false$ (by definition), the fact that $I$ is an
invariant implies that the post-state of any execution which reaches $v_{\rm e}$
must satisfy $\false$, and therefore the error location $v_{\rm e}$ is
unreachable.
Otherwise, $\cproof$ returns a new program path on which
our hypothesized invariant does not hold,
which we use
to refine our hypotheses.
In practice, one can perform $\cproof$ implicitly and lazily by maintaining a \emph{covering} (entailment)
relation~\cite{McMillan2006} over the nodes of the ART.

\begin{figure}[t]
\centering
    \begin{algorithmic}[1]
    \Function{\sf{ProvePath}}{$\pi = v_1,\ldots,v_n$}
        \Statex // \emph{Check if path is feasible (can reach error location $v_{\rm e}$)}.
        \State let $k$ be largest integer s.t. $\post(v_1,\ldots,v_k)$ is defined. \label{line:err1}
        \State $\emph{symbHeap} \gets \post(v_1,\ldots,v_k)$
        \Statex
        \State // \emph{If $\pi$ is memory-infeasible,} 
        \State // \emph{only compute spatial interpolants,}
        \State // \emph{as theory interpolants are not needed}
        \If {$S_k \models \false$, where $\emph{symbHeap} = \ldots,(S_k,v_k)$} \label{line:mem}
            \State let $k' \leq k$ be the largest integer s.t. $S_{k'} \not\models \false$
            \State $\emph{spInt} \gets \textsf{Spatial}((v_1,\ldots,v_{k'}), S_{k'})$
           \State\Return $\emph{spInt},(\false,v_{k'+1}),\ldots,(\false,v_n)$
        \EndIf \label{line:mem2}
        
        \State $\emph{refSymbHeap} \gets \refine(\emph{symbHeap},\false)$ 
        \If {$k = 0$ or \emph{refSymbHeap} is undefined}
            \State\Return no proof found  \label{line:err2}
        \EndIf
\\
        \Statex // \emph{Path infeasible -- construct proof}.
        \State $\emph{spInt} \gets \textsf{Spatial}((v_1,\ldots,v_k),\ltrue:\true)$  \label{line:pf1}
        \State $\emph{spTheoryInt} \gets \textsf{Refine}(\emph{spInt}, \false)$
        \If {\emph{spTheoryInt} is undefined}
            \State\Return $\emph{refSymbHeap},(\false,v_{k+1}),\ldots,(\false,v_n)$ \label{line:sym}
        \EndIf
        \State\Return $\emph{spTheoryInt},(\false,v_{k+1}),\ldots,(\false,v_n)$ \label{line:pf2}
    \EndFunction
    \end{algorithmic}
\caption{Pseudocode for \ppath.}
\label{alg:rpath}
\end{figure}

\paragraph{\ppath: Constructing a Proof of a Path}
Figure~\ref{alg:rpath} shows an algorithm
for computing a proof of a path $\pi$.
First, in lines~\ref{line:err1}-\ref{line:err2},
we check if $\pi$ is feasible (i.e., whether $\pi$
corresponds to a real program execution).
We do this by computing the strongest postconditions along $\pi$
(using $\post$)
and then attempting to strengthen the annotation (using \refine) with theory interpolants to prove that
\textit{false} is a postcondition of $\pi$.
If no such strengthening is found, we know that $\pi$ is feasible
(by Theorem~\ref{thm:refine_complete}).
Note that if $\pi$ is memory-infeasible (lines~\ref{line:mem}-\ref{line:mem2}),
then we only compute spatial interpolants along the memory-feasible
prefix of the path and return the result. This is because
when the path is memory-infeasible, we do not need data refinements
along the path to prove it cannot reach $v_{\rm e}$.

The function $\textsf{Spatial}((v_1,\ldots,v_n),P)$ takes a program path $\pi = v_1,\ldots,v_n$ and a \sep
formula $P$ and returns the path annotated with spatial interpolants w.r.t the postcondition $P$.
The function $\textsf{Refine}(\kappa,\varphi)$ takes an annotated program path $\kappa$
with $\sep$ formulas (from spatial interpolants) and a $\varphi \in \textsf{DFormula}$ and returns a
refined annotation of $\kappa$ that proves the postcondition $\varphi$ (using theory interpolants).

If the path $\pi$ is infeasible, we proceed by constructing spatial($\theory$) interpolants
for it (lines~\ref{line:pf1}-\ref{line:pf2}).
We use the function \textsf{Spatial} (Section~\ref{sec:spint}) to construct spatial interpolants,
which we then refine with theory interpolants using the function \textsf{Refine} (Section~\ref{sec:snint}).
Spatial path interpolants are computed with respect to the postcondition $\etrue : \true$,
indicating that we are looking for a memory safety proof.
Note that we might not be able to find theory interpolants if the spatial interpolants
computed are too weak and \emph{hide important data elements} 
(in which case, on line~\ref{line:sym}, we use the result of $\post$
as the spatial interpolants -- the strongest possible spatial interpolants).
To illustrate how this could happen, consider Figure~\ref{code:ref}: a modification
to our illustrative example from Figure~\ref{code:ex}.

\begin{figure}
\centering
\begin{minipage}{4.9cm}

\begin{lstlisting}
   node* x = null; int i = 2;
   while (i > 0)
     node* tmp = malloc(node);
     tmp->N = x; tmp->D = i;
     x = tmp; i--;
   i = 1;
P: while (x != null)
     if (isOdd(i))
       assert(isOdd(x->D))
     x = x->N; i++;
\end{lstlisting}
\end{minipage}
\caption{Refinement Example.}
\vspace{0.1in}
\label{code:ref}
\end{figure}


Here, a list of length 2 is constructed.
The second loop checks that nodes at odd positions in the linked
list have odd data elements.
Suppose \alg samples the path that goes through the 
first loop twice and enters the second loop arriving at the assertion.
Our spatial interpolation procedure will introduce a list segment $\ls(x,\nil)$
at location \texttt{P}. As a result, 
we cannot find a refinement that proves the assertion,
since using the list segment predicate definition from Section~\ref{sec:prelims} 
we cannot specify that the first element is odd and the second is even.
This is because  refinements must apply to all elements of $\ls$, 
and cannot mention specific positions.
In this case, we use the symbolic heaps as our spatial interpolants.
That is, we annotate location \texttt{P} with $x\mapsto [d_1',e'] * e' \mapsto [d_2',\nil]$.
Consequently, theory interpolants are able to refine this by specifying that $d_1'$ is odd.

\paragraph{\conj: Conjoining Proofs of Paths}
When a new annotated path $\kappa$ is computed,
we strengthen the proofs of all annotated
paths $\kappa'$ in $S$ that share a prefix with $\kappa$
using an operation $\conj(\kappa,\kappa')$, defined in the following.
(This is analogous to strengthening annotations of a path
in an ART -- all other paths sharing
a prefix with the strengthened path also get strengthened.)
Let $\kappa = (a_1,v_1),\ldots,(a_n,v_n)$
and $\kappa' = (a_1',v_1'),\ldots,(a_{m}',v_{m}')$.
Let $k$ be the largest integer such that
for all $j \leq k$, $v_j = v_j'$
(i.e., $k$ represents
the longest shared prefix of $\kappa$ and $\kappa'$).
$\conj$ returns a pair $(\ol{\kappa},\ol{\kappa}')$ consisting of the
strengthened annotations:
\begin{align*}
\ol{\kappa} &\gets (a_1 \land a_1', v_1),\ldots,(a_k \land a_k', v_k), (a_{k+1},v_{k+1}),\ldots,(a_n,v_n)\\
\ol{\kappa}' &\gets (a_1 \land a_1', v_1),\ldots,(a_k \land a_k', v_k), (a_{k+1}',v_{k+1}'),\ldots,(a_{m}',v_{m}')
\end{align*}
The issue here is that $\logic$ is not closed under logical conjunction
($\land$), 
since we do not allow
logical conjunction of spatial conjunctions, e.g., $(\ls(x,y)*\true) \land (\ls(y,z)*\true)$.
In practice, we heuristically under-approximate
the logical conjunction, using the strongest postcondition of the shared
prefix to guide the under-approximation.
Any under-approximation which over-approximates the strongest postcondition
(including the strongest postcondition itself) is sound, but overly strong
annotations may not generalize to a proof of the whole program.
Note that the above transformation maintains the invariant
that all paths in $S$ are annotated with valid Hoare triples.

%% file: appendix.tex
\section{Proofs}

In this section, we present proof sketches for the theorems appearing in this
paper.

\subsection{Proof of Theorem~\ref{thm:spint_sound}}
\begin{theorem}
  Let $S$ and $I'$ be \logic formulas and let $c$ be a command such that
  $\post(c,S) \models I'$.  Then
  \begin{enumerate}
  \item[(I)] $S \models \interp{S}{c}{I'}$
  \item[(II)] $\hoare{\interp{S}{c}{I'}}{c}{I'}$
  \end{enumerate}
\end{theorem}
\begin{proof}
  We prove one case to give intuition on why the theorem holds.

  Suppose \texttt{c} is an allocation statement \texttt{x := new($n$,$m$)}.

    Recall that we defined
    \[ \interp{S}{c}{I'} = \qex{x}{A} \]
    where \[\abduce{\post(c,S)}{\emptyset}{\qex{\vec{a},\vec{z}}{x \mapsto [\vec{a},\vec{z}]}}{I'}\]  Define
    \[ S' = \post(c,S) = S[x'/x] * x \mapsto [\vec{a},\vec{z}] \]
    for $x',\vec{a},\vec{z}$ fresh.

    First we show (I).  By the properties of bounded abduction, we have
    \[ S' = S[x'/x] * x \mapsto [\vec{a},\vec{z}] \models A * (\qex{\vec{a},\vec{z}}{x \mapsto [\vec{a},\vec{z}]}) \]
    from which we can see that
    \[ S[x'/x] \models A \models \qex{x}{A} \]
    and thus
    \[ S \models \qex{x}{A} = \interp{S}{c}{I'} \]

    Next we show (II).  We compute
    \begin{align*}
      \post(c,\qex{x}{A}) &= \qex{x',\vec{a},\vec{z}}{(\qex{x}{A})[x'/x] * x \mapsto [\vec{a},\vec{z}]}\\
      & \hspace*{10pt}\hfill \emph{where $x',\vec{a},\vec{z}$ are fresh.}\\
      & \equiv \qex{x',\vec{a},\vec{z}}{(\qex{x}{A}) * x \mapsto [\vec{a},\vec{z}]}\\
      & \equiv (\qex{x}{A}) * (\qex{\vec{a},\vec{z}}{x \mapsto [\vec{a},\vec{z}]})
    \end{align*}
    Since $S'$ is of the form $S[x'/x] * x \mapsto [\vec{a},\vec{z}]$ and $S' \models A
    * \true$, the only place where $x$ may appear in $A$ is in a disequality
    with some other allocated variable.  It follows that
    \[ (\qex{x}{A}) * (\qex{\vec{a},\vec{z}}{x \mapsto [\vec{a},\vec{z}]}) \equiv A * (\qex{\vec{a},\vec{z}}{x \mapsto [\vec{a},\vec{z}]}) \]
    By the properties of bounded abduction, we have
    \[ A * (\qex{\vec{a},\vec{z}}{x \mapsto [\vec{a},\vec{z}]}) \models I \]
    and thus
    \[ \post(c,\qex{x}{A}) \models I \]
    and finally
    \[ \hoare{\interp{S}{c}{I'}}{c}{I'} \]
\end{proof}

\subsection{Proof of Theorem~\ref{thm:refine_sound}}
\begin{theorem}[Soundness]
  Suppose that $\pi$ is a path and that $\zeta$ is a proof of the judgement
  $\cjudge{\hoare{\Pi:\Sigma}{$\pi$}{\Pi':\Sigma'}}{\mathcal{C}}$, and that
  $\sigma$ is a solution to $\mathcal{C}$.  Then $\zeta^\sigma$, the proof
  obtained by applying the substitution $\sigma$ to $\zeta$, is a (refined)
  separation logic proof of \[
  \hoare{(\Pi:\Sigma)^\sigma}{$\pi$}{(\Pi':\Sigma')^\sigma}\ .\]
\end{theorem}
\begin{proof}
  The proof proceeds by induction on $\zeta$.  We will give an illustrative
  example using an entailment judgement.

  Suppose that $\zeta$ is an entailment proof consisting of a single
  application of the \textsc{Predicate} rule:
  \begin{mathpar}
  \inferrule*[lab=Predicate]{
    \Pi \models \Pi' \\
  }{
    \Phi' \gets \Phi; \Psi_1' \gets \Psi_1 \land \Phi; \dotsi; \Psi_{|\vec{\tau}|}' \gets \Psi_{|\vec{\tau}|} \land \Phi
    \triangleright\\
    \Pi \land \Phi : Z(\vec{\tau},\vec{E}) \vdash \Pi' \land \Phi' : Z(\vec{\tau'},\vec{E})
  }
  \end{mathpar}
  (where $\tau_i = \lda{\vec{a}_i. \Psi_i}$ and $\tau_i' = \lda{\vec{a}_i. \Psi_i'}$).  Suppose that $\sigma$ is a solution to the constraint system
\[ \mathcal{C} = \Phi' \gets \Phi; \Psi_1' \gets \Psi_1 \land \Phi; \dotsi; \Psi_{|\vec{\tau}|}' \gets \Psi_{|\vec{\tau}|} \land \Phi \]

Since $\sigma$ is a solution to $\mathcal{C}$, we have that 
\[ \Phi'^\sigma \Leftarrow \Phi^\sigma \text{ and for all $i$, } \Psi_i'^\sigma \Leftarrow \Psi_i^\sigma \land \Phi^\sigma \]
and thus (noting that $\Pi \models \Pi'$)
\[ \Pi \land \Phi^\sigma \models \Pi' \land \Phi'^\sigma \text{ and for all $i$, } \Psi_i^\sigma \land \Pi \land \Phi^\sigma \models \Psi_i'^\sigma\]
It follows that $\zeta^\sigma$, given below, is a valid derivation:
  \begin{mathpar}
  \inferrule*[lab=Predicate]{
    \Pi \land \Phi'^\sigma \models \Pi' \land \Phi'^\sigma \\
    \Psi_1^\sigma \land \Pi \land \Phi^\sigma \models \Psi_1'\\
    \dotsi\\
    \Psi_n^\sigma \land \Pi \land \Phi^\sigma \models \Psi_n'^\sigma
  }{
    \Pi \land \Phi^\sigma : Z(\vec{\tau}_0^\sigma,\vec{E}) \vdash \Pi' \land \Phi'^\sigma : Z(\vec{\tau}_1^\sigma,\vec{E})
  }
  \end{mathpar}







\end{proof}

\subsection{Proof of Theorem~\ref{thm:refine_complete}}
\begin{theorem}[Completeness]
  Suppose that $\pi$ is a memory-safe path and $\zeta$ is the proof of the judgement
  \[\cjudge{\hoare{R_0(\vec{v}):\emp}{$\pi$}{R_1(\vec{v}) : \true}}{\mathcal{C}}\] obtained by symbolic execution.  If $\phi$ is a data formula such that 
  $\hoare{\textit{true}:\emp}{$\pi$}{\phi : \true}$
  holds, then there is a solution $\sigma$ to $\mathcal{C}$ such that
  $R_1^\sigma(\vec{v}) \Rightarrow \phi$.
\end{theorem}
\begin{proof}
  Consider that for each formula $\qs{X}{\Pi}{\Sigma}$ in a symbolic execution
  sequence, $\Sigma$ is a *-conjunction of (finitely many) points-to
  predicates.  The constraints we generate in this situation are the same as
  the ones that would be generated for a program which does not access the
  heap (but which has additional variables corresponding to data-typed heap
  fields).
\end{proof}

\section{Formalization of \logic and \sep}
\label{sec:logic}
In this section we present the full proof systems for \logic and \sep, as well
as the full set of constraint generation rules which we described in
Section~\ref{sec:snint}.
The syntax and semantics of \logic
formulas, in terms of stacks and heaps, are shown in Figures~\ref{fig:logic-full-syntax} and~\ref{fig:logic-full-semantics}.

\begin{figure*}[t]
\begin{minipage}[b]{0.45\linewidth}
\textbf{\emph{Syntax}}
\begin{align*}
x,y \in \textsf{HVar}& \hspace*{60pt}\text{Heap variables} \\
  a,b \in \dvar& \hspace*{60pt}\text{Data variables} \\
  A \in \textsf{DTerm} & \hspace*{60pt} \text{First-order term that evaluates to value in $\mathds{D}$}\\
  \phi \in \textsf{DFormula} & \hspace*{60pt} \text{Data formulas}\\
  Z \in \pred & \hspace*{60pt} \text{Recursive predicates}\\
  \theta \in \textsf{Refinement} &::= \lambda \vec{a}. \phi\\
  X \subseteq \textsf{Var}&::= x \mid a\\
  E,F \in \textsf{HTerm}&::= \nil \mid x\\
  \AE                  &::= A \ \mid E\\
  \Pi \in \textsf{Pure}&::= \etrue \mid E = E \mid E \neq E \mid \varphi \ \mid \Pi \land \Pi\\
  \Atom \in \textsf{Heaplet}&::= \true \mid \emp \mid E \mapsto [\vec{A},\vec{E}] \mid
Z(\vec{\theta},\vec{E})\\
  \Sigma \in \textsf{Spatial}&::= \Atom \mid \Atom * \Sigma\\
  P \in \logic&::= \qs{X}{\Pi}{\Sigma}
\end{align*}
\end{minipage}
\caption{Syntax of \logic formulas.}
\label{fig:logic-full-syntax}
\end{figure*}

\begin{figure*}[t]
\begin{minipage}[b]{0.45\linewidth}
\textbf{\emph{Semantic Domains}}
\begin{align*}
\textsf{Var} &= \textsf{HVar} + \dvar\\
\textsf{Val} &= \textsf{Loc} + \domain\\
\textsf{Stack} &= \textsf{Var} \rightarrow \textsf{Val}\\
\textsf{Heap} &= \textsf{Loc} \rightharpoonup_{\textsf{fin}} \rec\\
\rec &= \tuple{\mathds{N} \rightharpoonup_{\textsf{fin}} \domain, \mathds{N} \rightharpoonup_{\textsf{fin}} \textsf{Loc}}\\
\textsf{State} &= \textsf{Stack} \times \textsf{Heap}
\end{align*}
\end{minipage}

\begin{minipage}[b]{0.45\linewidth}
~\\

\textbf{\emph{Satisfaction Semantics}}
\begin{align*}
  s,h \models E = F &\iff \sem{E}(s) = \sem{F}(s)\\
  s,h \models E \neq F &\iff \sem{E}(s) \neq \sem{F}(s)\\
  s,h \models \varphi &\iff \sem{\varphi}(s)\\
  s,h \models \Pi_1 \land \Pi_2 &\iff (s,h \models \Pi_1) \text{ and } (s,h \models \Pi_2)\\
  s,h \models Z(\vec{\tau},\vec{E}) &\iff \exists P \in \mathit{cases}(Z(\vec{R},\vec{x})). s,h \models P[\vec{\tau}/\vec{R},\vec{E}/\vec{x}]\\
  s,h \models \emp &\iff \dom(h) = \emptyset\\
  s,h \models E \mapsto [\vec{A}, \vec{F}] & \iff \dom(h) = \{\sem{E}(s)\} \\ &\hspace*{1cm}\text{and } 
        h(\sem{E}(s)) =  &\\
         &\hspace*{1.25cm} \tuple{\{i \mapsto \sem{A_i}(s) | i \in [1,|\vec{A}|]\}, \{i \mapsto \sem{F_i}(s) | i \in [1,|\vec{F}|]\}}\\
  s,h \models \Sigma * \Sigma' &\iff \text{there exists } h_0, h_1 \text{ s.t. }  h_0 \uplus h_1 = h \\ &\hspace*{1cm} \text{and } (s,h_0 \models \Sigma) \text{ and } (s,h_1 \models \Sigma')\\
  s,h \models \Pi \colon \Sigma &\iff (s,h \models \Pi) \text{ and } (s,h \models \Sigma)\\
  s,h \models \qs{X}{\Pi}{\Sigma} &\iff \text{there exists } \overline{s} : X \rightarrow \textsf{Val} \text{ s.t. }
  s\oplus \overline{s},h \models \Pi:\Sigma
\end{align*}
\end{minipage}
~\\
Note that we model records (\rec) as two finite maps representing data fields and heap fields.\\
We use $\uplus$ to denote union of functions with disjoint domains, and $\oplus$ to
denote overriding union of functions.

\caption{Stack/heap semantics of \logic formulas.}
\label{fig:logic-full-semantics}
\end{figure*}

\begin{figure*}[t]
  \footnotesize
\figsep{Entailment rules}
\begin{mathpar}
  \inferrule*[lab=Empty]{\Pi \models \Pi'}{ \Pi : \emp \vdash \Pi' : \emp}

  \inferrule[$\exists$-left]{
    P[x'/x] \vdash Q
  }{
    \qex{x}{P} \vdash Q
  }

  \inferrule[$\exists$-right]{
    P \vdash Q[\mbox{\AE}/x]
  }{
    P \vdash \qex{x}{Q}
  }

  \inferrule*[lab=Predicate,right={\rm\begin{minipage}{2.85cm}
        Where $\tau_i = \lda{\vec{a}_i. \phi_i}$\\
        and $\tau_i' = \lda{\vec{a}_i. \phi_i'}$
      \end{minipage}}]{
    \Pi \models \Pi' \\
    \phi_1 \land \Pi \models \phi_1'\\
    \dotsi\\
    \phi_n \land \Pi \models \phi_n'
  }{
    \Pi : Z(\vec{\tau},\vec{E}) \vdash \Pi' : Z(\vec{\tau}',\vec{E})
  }

  \inferrule*[lab=True]{\Pi \models \Pi'}{
    \Pi : \Sigma \vdash \Pi' : \true
  }

  \inferrule[Points-to]{\Pi \models \Pi'}{
    \Pi : E \mapsto [\vec{A},\vec{F}] \vdash \Pi' : E \mapsto [\vec{A},\vec{F}]
  }

  \inferrule[Star]{
    \Pi : \Sigma_0 \vdash \Pi' : \Sigma_0' \\
    \Pi : \Sigma_1 \vdash \Pi' : \Sigma_1'
  }{
    \Pi : \Sigma_0 * \Sigma_1  \vdash \Pi' : \Sigma_0' * \Sigma_1'
  }

  \inferrule[Substitution]{
    \Pi[E/x] : \Sigma[E/x] \vdash \Pi'[E/x] : \Sigma'[E/x] \\
    \Pi \models x = E
  }{
    \Pi : \Sigma \vdash \Pi' : \Sigma'
  }

  \inferrule[\nil-not-Lval]{
    \Pi \land E \neq \nil : \Sigma * E \mapsto [\vec{A},\vec{F}] \vdash \Pi' : \Sigma'
  }{
    \Pi : \Sigma * E \mapsto [\vec{A},\vec{F}] \vdash \Pi' : \Sigma'
  }

    \inferrule[*-Partial]{
    \Pi \land E \neq F : \Sigma * E \mapsto [\vec{A},\vec{E}] * F \mapsto [\vec{B},\vec{F}] \vdash \Pi' : \Sigma'
  }{
    \Pi : \Sigma * E \mapsto [\vec{A},\vec{E}] * F \mapsto [\vec{B},\vec{F}] \vdash \Pi' : \Sigma'
  }

  \inferrule*[lab=Fold,right={\rm $P \in \mathit{cases}(Z(\vec{R},\vec{x}))$}]{
    \Pi : \Sigma \vdash \Pi' : \Sigma' * P[\vec{\tau}/\vec{R},\vec{E}/\vec{x}]
  }{
    \Pi : \Sigma \vdash \Pi' : \Sigma' * Z(\vec{\tau},\vec{E})
  }

  \inferrule*[lab=Unfold,right={\rm $\{P_1,\dotsc,P_n\} = \mathit{cases}(Z(\vec{R},\vec{x}))$}]{
    \Pi : \Sigma * P_1[\vec{\tau}/\vec{R},\vec{E}/\vec{x}] \vdash \Pi' : \Sigma'\\
    \dotsi\\
    \Pi : \Sigma * P_n[\vec{\tau}/\vec{R},\vec{E}/\vec{x}] \vdash \Pi' : \Sigma'
  }{
    \Pi : \Sigma * Z(\vec{\tau},\vec{E}) \vdash \Pi' : \Sigma'
  }
\end{mathpar}
\caption{\logic Proof System}
\label{fig:qsep-ent}
\end{figure*}

\begin{figure*}[t]
\figsep{Execution rules}
\begin{mathpar}
  \inferrule[Assign]{ }{
    \hoare{\Pi \colon \Sigma}
          {x := \AE}
          {\qex{x'}{\Pi[x'/x] \land x = \mbox{\AE}[x'/x] : \Sigma[x'/x]}}
  }

  \inferrule[Assume]{ }{
    \hoare{\Pi \colon \Sigma}{assume($\Pi'$)}{\Pi \land \Pi' : \Sigma}
  }

  \inferrule[Sequence]{
    \hoare{P}{$\pi_0$}{O}\\
    \hoare{O}{$\pi_1$}{Q}
  }{
    \hoare{P}{$\pi_0;\pi_1$}{Q}
  }

  \inferrule[Data-Store]{
    \Pi \colon \Sigma \vdash \Pi' : \Sigma' * x \mapsto [\vec{A},\vec{E}]
  }{
    \hoare{\Pi \colon \Sigma}
          {x->D$_i$ := A}
          {\Pi' : \Sigma' * x \mapsto [\vec{A}[A/A_i],\vec{E}]}
  }

  \inferrule[Heap-Store]{
    \Pi \colon \Sigma \vdash \Pi' : \Sigma' * x \mapsto [\vec{A},\vec{E}]
  }{
    \hoare{\Pi \colon \Sigma}{x->N$_i$ := E}{\Pi' : \Sigma' * x \mapsto [\vec{A},\vec{E}[E/E_i]]}
  }

  \inferrule[Data-Load]{
    \Pi \colon \Sigma \vdash \Pi' : \Sigma' * x \mapsto [\vec{A},\vec{E}]
  }{
    \hoare{\Pi \colon \Sigma}
          {y := x->D$_i$}
          {\qs{y'}
            {\Pi'[y'/y] \land y = A_i[y'/y]}
            { (\Sigma' * x \mapsto [\vec{A},\vec{E}])[y'/y]}}
  }

  \inferrule[Free]{
    \Pi \colon \Sigma \vdash \Pi' : \Sigma' * x \mapsto [\vec{A},\vec{E}]
  }{
    \hoare{\Pi \colon \Sigma}{free(x)}{\Pi' : \Sigma'}
  }

  \inferrule[Heap-Load]{
    \Pi \colon \Sigma \vdash \Pi' : \Sigma' * x \mapsto [\vec{A},\vec{E}]
  }{
    \hoare{\Pi \colon \Sigma}
          {y := x->N$_i$}
          {\qs{y'}
            {\Pi'[y'/y] \land y = E_i[y'/y]}
            { (\Sigma' * x \mapsto [\vec{A},\vec{E}])[y'/y]}}
  }

  \inferrule[Consequence]{
    P' \vdash P\\
    \hoare{P}{c}{Q}\\
    Q \vdash Q'
  }{
    \hoare{P'}{c}{Q'}
  }

  \inferrule[Alloc]{ }{
    \hoare{\Pi \colon \Sigma}
          {x := new(n,m)}
          {\qs{x',\vec{a},\vec{y}}{\Pi[x'/x] }{ \Sigma[x'/x] * x \mapsto [\vec{a},\vec{y}]}}
  }

  \inferrule*[lab=Exists,right={\rm $x \notin \Vars(Q) \cup \Vars(c)$}]{
    \hoare{P}{c}{Q}
  }{
    \hoare{\qex{x}{P}}{c}{Q}
  }
\end{mathpar}
\caption{\logic proof system.}
\label{fig:qsep-exec}
\end{figure*}

\begin{figure*}[t]
\begin{mathpar}
  \small
  \inferrule*[lab=Predicate]{
    \Pi \models \Pi'
  }{
    \Pi : Z(\vec{E}) \vdash \Pi' : Z(\vec{E})
  }

  \inferrule*[lab=Fold,right={\rm $P \in \mathit{cases}(Z(\vec{R},\vec{x}))$}]{
    \Pi : \Sigma \vdash \Pi' : \Sigma' * P[\vec{E}/\vec{x}]
  }{
    \Pi : \Sigma \vdash \Pi' : \Sigma' * Z(\vec{E})
  }

  \inferrule*[lab=Unfold,right={\rm $\{P_1,\dotsc,P_n\} = \mathit{cases}(Z(\vec{R},\vec{x}))$}]{
    \Pi : \Sigma * P_1[\vec{E}/\vec{x}] \vdash \Pi' : \Sigma'\\\\
    \dotsi\\\\
    \Pi : \Sigma * P_n[\vec{E}/\vec{x}] \vdash \Pi' : \Sigma'
  }{
    \Pi : \Sigma * Z(\vec{E}) \vdash \Pi' : \Sigma'
  }
\end{mathpar}
\caption{\sep proof system.  All other entailment and execution rules are as
  in Figure~\ref{fig:qsep-ent}.}
\label{fig:symb}
\end{figure*}

\begin{figure*}[t]
\figsep{Entailment rules}
\vspace{-.15in}
\begin{mathpar}
  \footnotesize
  \inferrule[Empty]{
    \Pi \models \Pi'
  }{
    \cjudge{\Pi \land \Phi : \emp \vdash \Pi' \land \Phi' : \emp}
           {\Phi' \gets \Phi}
  }

  \inferrule[True]{\Pi \models \Pi'}{
    \cjudge{\Pi \land \Phi : \Sigma \vdash \Pi' \land \Phi' : \true}
           {\Phi' \gets \Phi}
  }

  \inferrule[Inconsistent]{
    \Pi \models \textit{false}
  }{
    \cjudge{\Pi \land \Phi : \Sigma \vdash \Pi' \land \Phi' : \Sigma'}
           { [] }
  }

  \inferrule[$\exists$-left]{
    \cjudge{P[x'/x] \vdash Q}{\mathcal{C}}
  }{
    \cjudge{\qex{x}{P} \vdash Q}{\mathcal{C}}
  }

  \inferrule[Substitution]{
    \cjudge{\Pi[E/x] \land \Phi : \Sigma[E/x] \vdash \Pi'[E/x] \land \Phi' : \Sigma'[E/x]}
           {\mathcal{C}}\\
    \Pi \models x = E
  }{
    \cjudge{\Pi \land \Phi : \Sigma \vdash \Pi' \land \Phi' : \Sigma'}
           {\mathcal{C}}
  }

  \inferrule[\nil-not-Lval]{
    \cjudge{\Pi \land \Phi \land E \neq \nil : \Sigma * E \mapsto [\vec{A},\vec{F}] \vdash \Pi' \land \Phi' : \Sigma'}
           {\mathcal{C}}
  }{
    \cjudge{\Pi \land \Phi : \Sigma * E \mapsto [\vec{A},\vec{F}] \vdash \Pi' \land \Phi' : \Sigma'}
           {\mathcal{C}}
  }

    \inferrule[$\exists$-right]{
    \cjudge{P \vdash Q[\mbox{\AE}/x]}{\mathcal{C}}
  }{
    \cjudge{P \vdash \qex{x}{Q}}{\mathcal{C}}
  }

  \inferrule[*-Partial]{
    \cjudge{\Pi \land E \neq F \land \Phi : \Sigma * E \mapsto [\vec{A},\vec{E}] * F \mapsto [\vec{B},\vec{F}] \vdash \Pi' \land \Phi' : \Sigma'}
           {\mathcal{C}}
  }{
    \cjudge{\Pi \land \Phi : \Sigma * E \mapsto [\vec{A},\vec{E}] * F \mapsto [\vec{B},\vec{F}] \vdash \Pi' \land \Phi' : \Sigma'}
           {\mathcal{C}}
  }

  \inferrule[Star]{
    \cjudge{\Pi \land \Phi : \Sigma_0 \vdash \Pi' \land \Phi' : \Sigma_0'}
           { \mathcal{C}_0} \\
    \cjudge{\Pi \land \Phi : \Sigma_1 \vdash \Pi' \land \Phi' : \Sigma_1'}
           { \mathcal{C}_1 }  
  }{
    \cjudge{\Pi \land \Phi : \Sigma_0 * \Sigma_1  \vdash \Pi' \land \Phi' : \Sigma_0' * \Sigma_1'}
           {\mathcal{C}_0;\mathcal{C}_1}
  }

  \inferrule[Points-to]{ \Pi \models \Pi' } {
    \cjudge{\Pi \land \Phi : E \mapsto [\vec{A},\vec{F}] \vdash \Pi' \land \Phi' : E \mapsto [\vec{A},\vec{F}]}
           { \Phi' \gets \Phi }
  }

  \inferrule*[lab=Fold,right={\rm $P \in \mathit{cases}(Z(\vec{R},\vec{x}))$}]{
    \cjudge{\Pi : \Sigma \vdash \Pi' : \Sigma' * P[\vec{\tau}/\vec{R},\vec{E}/\vec{x}]}
           {\mathcal{C}}
  }{
    \cjudge{\Pi : \Sigma \vdash \Pi' : \Sigma' * Z(\vec{\tau},\vec{E}) }
           {\mathcal{C}}
  }

  \inferrule*[lab=Unfold,right={\rm $\{P_1,\dotsc,P_n\} = \mathit{cases}(Z(\vec{R},\vec{x}))$}]{
    \cjudge{\Pi : \Sigma * P_1[\vec{\tau}/\vec{R},\vec{E}/\vec{x}] \vdash \Pi' : \Sigma' }
           {\mathcal{C}_1}\\
           \dotsi\\
    \cjudge{\Pi : \Sigma * P_n[\vec{\tau}/\vec{R},\vec{E}/\vec{x}] \vdash \Pi' : \Sigma'}
           {\mathcal{C}_n}
  }{
    \cjudge{\Pi : \Sigma * Z(\vec{\tau},\vec{E}) \vdash \Pi' : \Sigma'}
           {\mathcal{C}_1; \dotsc{;}\, \mathcal{C}_n}
  }

  \inferrule*[lab=Predicate,right={\rm\begin{minipage}{2.85cm}
        Where $\tau_i = \lda{\vec{a}_i. \Psi_i}$\\
        and $\tau_i' = \lda{\vec{a}_i. \Psi_i'}$
      \end{minipage}}]{
    \Pi \models \Pi' \\
  }{
    \Phi' \gets \Phi; \Psi_1' \gets \Psi_1 \land \Phi; \dotsc{;}\, \Psi_{|\vec{\tau}|}' \gets \Psi_{|\vec{\tau}|}' \land \Phi
    \triangleright
    \Pi \land \Phi : Z(\vec{\tau},\vec{E}) \vdash \Pi' \land \Phi' : Z(\vec{\tau'},\vec{E})
  }
\end{mathpar}
\caption{Constraint Generation: Entailment Rules}
\label{fig:constraint-ent}
\end{figure*}

\begin{figure*}[t]
\figsep{Execution rules}
\vspace{-.15in}
\begin{mathpar}
  \footnotesize
  \inferrule[Data-Assume]{
    \cjudge{P \land \phi \vdash Q}
           {\mathcal{C}}
  }{
    \cjudge{\hoare{P}
                  {assume($\phi$)}
                  {Q}}
           {\mathcal{C}}
  }

  \inferrule[Free]{
    \cjudge{P \vdash \Pi \land \Phi : \Sigma * x \mapsto [\vec{A},\vec{E}]}
           {\mathcal{C}}
  }{
    \cjudge{\hoare{P}{free(x)}{\Pi \land \Phi : \Sigma}}
           {\mathcal{C}}
  }

  \inferrule[Sequence]{
    \cjudge{\hoare{P}
                  {$\pi_0$}
                  {\widehat{O}}}
           {\mathcal{C}_0}\\
    \cjudge{\hoare{\widehat{O}}
                  {$\pi_1$}
                  {Q}}
           {\mathcal{C}_1}
  }{
    \cjudge{\hoare{P}
                  {$\pi_0;\pi_1$}
                  {Q}}
           {\mathcal{C}_0;\mathcal{C}_1}
  }

  \inferrule[Data-Load]{
    \cjudge{P \vdash \qex{X}{\Pi \land \widehat{\Phi} : \widehat{\Sigma} * x \mapsto [\vec{A},\vec{E}]}}
           {\mathcal{C}_0}\\\\
    \cjudge{\qex{X,a'}{\Pi[a'/a] \land \widehat{\Phi}[a'/a] \land a = A_i[a'/a] \colon (\widehat{\Sigma} * x \mapsto [\vec{A},\vec{E}])[a'/a]} \vdash Q}
           {\mathcal{C}_1}         
  }{
    \cjudge{\hoare{P}
                  {a := x->D$_i$}
                  {Q}}
           {\mathcal{C}_0; \mathcal{C}_1 }
  }

  \inferrule[Data-Assign]{
    \cjudge{\qex{a'}{\Pi \land \Phi[a'/a] \land a = A[a'/a] : \Sigma[a'/a] \vdash Q}}
           {\mathcal{C}}
  }{
    \cjudge{\hoare{\Pi \land \Phi \colon \Sigma}{a := A}{Q}}
           { \mathcal{C} }
  }

  \inferrule[Data-Store]{
    \cjudge{P \vdash \qex{X}{\Pi \land \widehat{\Phi} : \widehat{\Sigma} * x \mapsto [\vec{A},\vec{E}]}}
           {\mathcal{C}_0}\\\\
    \cjudge{\qex{X,a'}{\Pi \land \widehat{\Phi} \land a' = A : \widehat{\Sigma} * x \mapsto [\vec{A}[a'/A_i],\vec{E}]} \vdash Q}
           {\mathcal{C}_1}
  }{
    \cjudge{\hoare{P}{x->D$_i$ := A}{Q}}
           {\mathcal{C}_0; \mathcal{C}_1}
  }

  \inferrule[Alloc]{
    \cjudge{\qex{x',\vec{a},\vec{x}}{\Pi[x'/x] \land \Phi : \Sigma[x'/x] * x \mapsto [\vec{a},\vec{x}]} \vdash Q}
           {\mathcal{C}}
  }{
    \cjudge{\hoare{\Pi \land \Phi \colon \Sigma}
                  {x := new($n$,$m$)}
                  {Q}}
           {\mathcal{C}}
  }

  \inferrule[Heap-Store]{
    \cjudge{\Pi \colon \Sigma \vdash \Pi' : \Sigma' * x \mapsto [\vec{A},\vec{E}]}
           {\mathcal{C}}
  }{
    \cjudge{\hoare{\Pi \colon \Sigma}{x->N$_i$ := E}{\Pi' : \Sigma' * x \mapsto [\vec{A},\vec{E}[E/E_i]]}}
           {\mathcal{C}}
  }

  \inferrule[Consequence]{
    \cjudge{P' \vdash \widehat{P}}{\mathcal{C}_1}\\
    \cjudge{\hoare{\widehat{P}}{c}{\widehat{Q}}}{\mathcal{C}_2}\\
    \cjudge{\widehat{Q} \vdash Q'}{\mathcal{C}_3}
  }{
    \cjudge{\hoare{P'}{c}{Q'}}{\mathcal{C}_1;\mathcal{C}_2;\mathcal{C}_3}
  }

  \inferrule[Heap-Load]{
    \cjudge{\Pi \colon \Sigma \vdash \Pi' : \Sigma' * x \mapsto [\vec{A},\vec{E}]}
           {\mathcal{C}}
  }{
    \cjudge{\hoare{\Pi \colon \Sigma}
                  {y := x->N$_i$}
                  {\qs{y'}
                    {\Pi'[y'/y] \land y = E_i[y'/y]}
                    { (\Sigma' * x \mapsto [\vec{A},\vec{E}])[y'/y]}}}
           {\mathcal{C}}
  }
\end{mathpar}
\caption{Constraint Generation: Execution Rules}
\label{fig:constraint-exec}
\end{figure*}

\section{Spatial interpolation for assumptions}
\label{sec:assum}
In the spatial interpolation rules for \texttt{assume} presented in
Section~\ref{sec:spint}, we encountered the following problem: we have an
equality or disequality assertion $\Pi$, a symbolic heap $S$, and a formula
$M$ such that $S \land \Pi \vdash M$, and we need to compute a formula $M'$
such that $S \vdash M'$ and $M' \land \Pi \vdash M$.  Moreover, we wish $M'$
to be as weak as possible (i.e., $M'$ should be ``close to $M$'' rather than
``close to $S$'').

In this section, we will define a recursive procedure $\textsf{pitp}(S,\Pi,M)$
which takes as input a symbolic heap $S$, an equality or disequality formula
$\Pi$, and a $\sep$ formula $M$ such that $S \land \Pi \vdash M$ and computes
a formula $M' = \textsf{pitp}(S,\Pi,M)$ such that $S \vdash M'$ and $M' \land
\Pi \vdash M$.  We will assume that $S = \Pi_S : \Sigma_S$ is saturated in the
sense that for any assertion $\Pi_0$, if $S \vdash \Pi_0$ then $\Pi_S \vdash
\Pi_0$, and that $S \land \Pi$ is satisfiable.

We observe that if $M$ has existentially quantified variables, they can
be instantiated using the proof of $S \land \Pi \vdash M$.  Thus we may assume
that $M$ is quantifier-free, and write $M$ as
  \[ M = \Pi_M : H_1 * \dotsi * H_n \]
The proof of $S \land \Pi \vdash M$ also induces an $n$-colouring on $S$ which
colours each points-to predicate in $S$ with the index $i$ of the
corresponding heaplet $H_i$ (cf. step 1 of the bounded abduction procedure
presented in Section~\ref{sec:abduction}).  We may thus write $S$ as follows:
\[ S = \Pi_S : \coloured{\Sigma_1}{1} * \dotsi * \coloured{\Sigma_n}{n} \]
(such that for each $i$, $\Pi_S \land \Pi : \Sigma_i \vdash \Pi_M : H_i $).

We will compute a pure formula $\Pi_M'$ such that $\Pi_S \vdash \Pi_M'$ and
$\Pi_M' \land \Pi \vdash \Pi_M$ and for each $i$ we will compute a formula
$P_i$ such that $\Pi_S : \Sigma_i \vdash P_i$ and $P_i \land \Pi \vdash H_i$.
We then take $\textsf{pitp}(S,\Pi,M)$ to be $\Pi_M' : P_1 * \dotsi * P_n$.

First, we show how to compute $\Pi_M'$.  Note that note that since $S$ is
saturated, the fact that $S \land \Pi \vdash M$ implies $\Pi_S \land \Pi
\vdash \Pi_M$.  We will assume that $\Pi_M$ consists of a single equality or
disequality: the procedure can be extended to arbitrary conjunction by
applying it separately for each conjunct and conjoining the results.  If $\Pi_S
\vdash \Pi_M$ then we simply take $\Pi_M'$ to be $\Pi_M$.  Otherwise, assume
$w,x,y,z$ are such that $\Pi$ is an equality or disequality $w = x$ / $w \neq
x$ and $\Pi_M$ is an equality or disequality $y = z$ / $y \neq z$.  Since
$\Pi_S \land \Pi \vdash \Pi_M$ and $\Pi_S \not\vdash \Pi_M$, there is some
$y',z' \in \{w,x\}$ such that $\Pi_S \vdash y = y' \land z = z'$ (to see why,
consider each of the four cases for $\Pi_S$ and $\Pi_M$).  We take $\Pi_M'$ to
be $y = y' \land z = z'$.

Finally, we show how to compute $P_i$ (for each $i \in [1,n]$).  If $\Pi_S :
\Sigma_i \vdash H_i$, then we simply take $P_i$ to be $H_i$.  Otherwise,
suppose that $H_i$ is $Z(\vec{E})$ for some predicate $Z$ and vector of heap
terms $\vec{E}$ (the case that $H_i$ is a points-to predicate is essentially a
special case).  First, we attempt to find a vector of heap terms $\vec{E'}$
such that $\Pi_S : \Sigma_i \vdash Z(\vec{E'})$ and $\Pi_S \land \Pi \vdash
E_j = E_j'$ for each $j$ (noting that there are finitely many such $\vec{E'}$
to choose from).  If we succeed, we may take $P_i$ to be $\Pi' : Z(\vec{E'})$,
where $\Pi' = \textsf{pitp}(S, \Pi, \bigwedge_j E_j = E_j')$.  If we fail,
then since $\Pi_S \land \Pi : \Sigma_i \vdash Z(\vec{E})$, there is some $Q
\in \mathit{cases}(Z(\vec{R},\vec{x}))$ such that $\Pi_S \land \Pi : \Sigma_i \vdash
\ul{Q}[\vec{E}/\vec{x}]$.  We may take $P_i$ to be $\textsf{pitp}(\Pi_S : \Sigma_i,
\Pi, Q[\vec{E}/\vec{x}])$.